\documentclass[%
  reprint,
  amsmath,amssymb,
  aps,
  floatfix,
  superscriptaddress
]{revtex4-2} 

\usepackage[T1]{fontenc}
\usepackage[utf8]{inputenc}
\usepackage{graphicx}
\graphicspath{{media/}}
\usepackage{bm}
\usepackage{color}
\usepackage{dcolumn}
\usepackage{booktabs}
\usepackage{hyperref}
\usepackage{url}
\usepackage{microtype}
\usepackage{comment}
\usepackage{placeins}
\usepackage{float}

\begin{document}

\title{Probing Neutrino-Energy Reconstruction with New Superscaling-Based Observables}

\author{Lounès Amziane }
\email{lounes.amziane@unige.ch}
\affiliation{Université de Genève, Section de Physique, DPNC, 1205 Genève, Switzerland}

\author{Soniya Samani}
\affiliation{Université de Genève, Section de Physique, DPNC, 1205 Genève, Switzerland}

\author{Lorenzo Giannessi}
\affiliation{Université de Genève, Section de Physique, DPNC, 1205 Genève, Switzerland}
\affiliation{Johannes Gutenberg University Mainz, 55128 Mainz, Germany}

\author{Federico Sánchez }
\affiliation{Université de Genève, Section de Physique, DPNC, 1205 Genève, Switzerland}
\date{\today}

\begin{abstract}
In this study, we explore the superscaling variable \(\psi'\) within the context of neutrino--nucleus charged-current quasielastic interactions. Building on the neutrino-oriented definition of the superscaling variable \(\psi'\), we reparametrize it in terms of the neutrino-energy reconstruction bias, thereby establishing a direct conceptual and analytical connection between the superscaling formalism and neutrino-energy reconstruction. This approach allows the hadronic part of the reconstructed \(\psi'\) to be expanded around the quasielastic point, yielding new observables that depend only on the outgoing-muon kinematics. At the same time, the resulting distributions remain sensitive to the underlying nuclear dynamics and vary across nuclear models. Among them, a newly identified quantity \(L_1\), constructed
solely from the outgoing-muon kinematics, provides a nuclear-physics constraint on the neutrino-energy reconstruction bias. This observable largely inherits the scaling properties of \(\psi'\) across different neutrino fluxes and nuclear targets. The nuclear-model dependence is investigated using ND280-like $\nu_\mu$ and $\bar{\nu}_\mu$ Monte Carlo samples on carbon, generated with several nuclear models implemented in NEUT, NuWro, and GENIE, while the stability across experimental conditions is tested using additional carbon- and argon-based flux--target configurations.

\end{abstract}

\maketitle

\section{Introduction}
\label{sec:intro}

Precise neutrino--nucleus interaction modeling is increasingly important for oscillation experiments, where the reduction of statistical uncertainties puts nuclear effects among the leading systematic limitations \cite{PhysRevD.111.093010, Ankowski:2017systematics}. Superscaling in inclusive electron scattering provides a well-motivated framework to organize these effects. The superscaling variable $\psi'(\omega,|\vec q|)$ maps charged current quasi-elastic (CC1p1h, following the modern nomenclature) kinematics onto a dimensionless coordinate for which the reduced quasielastic response exhibits approximate scaling with momentum transfer and nuclear species \cite{Donnelly1999}. Its success in inclusive electron scattering therefore motivates analogous descriptions of neutrino–nucleus interactions. In neutrino physics, the superscaling framework has been developed along two main directions. On the one hand, the SuSAv2 approach uses superscaling information from inclusive electron scattering to construct phenomenological  predictions for neutrino--nucleus interactions and corresponding implementations in neutrino event generators~\cite{Megias:2013qcw,Dolan2020}, comparing model predictions to data at the level of differential cross sections rather than reconstructing $\psi'$ directly from neutrino events. On the other hand, reconstructed versions of $\psi'$ have also been proposed and studied directly as experimental observables in neutrino scattering, demonstrating that superscaling variables can be accessed in practice from neutrino data \cite{Douqa2024}.

Following the approach introduced in Ref.~\cite{Douqa2024}, we extend this construction to a measurable quantity in neutrino experiments by reconstructing the incoming neutrino energy from the outgoing proton kinematics. In this form, $\psi'$ acquires an explicit hadronic dependence, together with a reconstruction parameter, which can be interpreted as an effective energy shift that includes the average removal energy and a correction associated with nuclear rescattering. However, this reconstructed form of \(\psi'\) depends explicitly on the final-state proton kinematics and on an effective nuclear-energy parameter, limiting its applicability when the hadronic system is not fully reconstructed.

The central idea of this work is to express $\psi'$ in terms of the neutrino-energy reconstruction bias, denoted $\Delta E_\nu$, defined as the difference between the true and quasielastic reconstructed neutrino energies, and to perform a Taylor expansion around $\Delta E_\nu=0$. In this expansion, the hadronic dependence is encoded entirely in $\Delta E_\nu$, while the expansion coefficients, denoted $L_n$, depend only on the outgoing-muon kinematics. The leading coefficients, in particular $L_1$ and $L_2$, therefore define a new class of muon-only observables. These observables can be applied to both neutrino and antineutrino samples, including those collected at water-Cherenkov far detectors such as Super-Kamiokande and Hyper-Kamiokande~\cite{FUKUDA2003418,ProtoCollaboration2018}. They also have a clear interpretation in terms of neutrino-energy reconstruction and exhibit an approximately stable behavior across different neutrino fluxes and nuclear targets. A major result of this construction is the derivation of a new muon-only observable that provides an event-by-event, target-dependent upper bound on the magnitude of the neutrino-energy reconstruction bias, $|\Delta E_\nu|$, assuming the interaction lies within the quasielastic superscaling region $\lvert \psi' \rvert < 1$.

We investigate the sensitivity of the proposed observables to nuclear-model assumptions by comparing several theoretical descriptions within the \textsc{NUISANCE} framework~\cite{Stowell:2016jfr}, using ND280 $\nu_{\mu}$ and $\bar{\nu}_{\mu}$ CC1p1h samples on a carbon target simulated with a variety of neutrino event generators. 
Throughout this work, we restrict our study to the theoretical level, using truth-level Monte Carlo samples of CC1p1h events. The impact of experimental effects such as background contamination from non-CC1p1h channels, detector acceptance, and measurement systematics is left for future experimental studies.

The remainder of this paper is organized as follows. Section~\ref{sec:energy reco} establishes the reconstruction formalism, defining the energy reconstruction (RE) and lepton-only quasielastic (LQE) schemes. In Section~\ref{sec:L1_weak_EnuE}, we rewrite $\psi'$ in terms of the neutrino-energy reconstruction bias, and use its Taylor expansion to derive new muon-only observables and the associated reconstruction bounds. Section~\ref{sec:model_comparison} investigates the sensitivity of the reconstructed $\psi'$ and the derived new observables to different nuclear-model assumptions, assessing their sensitivity to differences among generator and model configurations. Section~\ref{sec:exp_outlook} investigates the behavior of these observables across different neutrino fluxes and nuclear targets, assessing the extent to which they remain stable under variations in the flux--target configuration. Finally, Section~\ref{sec:conclusions} summarizes our findings and discusses the potential of these new observables as constraints on neutrino-energy reconstruction systematics in oscillation analyses.

\section{Neutrino and antineutrino energy reconstruction}
\label{sec:energy reco}

In charged-current one-particle--one-hole (CC1p1h) interactions, the primary weak interaction is
\[
\nu_\mu+n\rightarrow\mu^-+p,
\]
or, for antineutrinos,
\[
\bar{\nu}_\mu+p\rightarrow\mu^++n,
\]
where the nucleons are initially bound inside the nucleus. Two standard neutrino-energy reconstruction strategies can therefore be considered: a hadronic reconstruction using the outgoing muon ($\mu^\pm$) and nucleon ($p$ or $n$) kinematics, and a quasielastic reconstruction based only on the outgoing muon kinematics. Since, as we will see in later sections, the reconstruction of $\psi'$ and of the neutrino-energy reconstruction bias relies on both approaches, we first introduce a unified framework that consistently accounts for the removal-energy convention. Unless otherwise stated, the illustrative distributions shown in this section are obtained from representative CC1p1h samples generated with NEUT v6.1.4 using the local Fermi gas (LFG) model~\cite{Hayato2021,Bourguille2021,PhysRevD.88.113007}.

In neutrino experiments, the neutrino energy is not measured directly but must be inferred from the kinematics of the outgoing particles. The outgoing muon is generally observable, whereas the outgoing-nucleon energy is measurable only in certain detector configurations. A clear example is antineutrino scattering, where the outgoing neutron is typically not reconstructed, although recent developments~\cite{Manly:2025pfm} suggest that this may become possible in the future.

As an example, in T2K, the outgoing-proton information required for hadronic energy reconstruction is accessible at the near detector, ND280, but not at the water-Cherenkov far detector, Super-Kamiokande, where the neutrino energy is reconstructed from the outgoing-lepton kinematics. This motivates the introduction of two neutrino-energy reconstruction methods. The first uses both the outgoing-muon and outgoing-nucleon kinematics together with an assumed removal energy and is referred to as energy reconstruction (RE). The second uses only the outgoing-muon kinematics under the quasielastic target-at-rest hypothesis and is referred to as leptonic quasielastic reconstruction (LQE).

\subsection{Event-level removal energy}
\label{sec:nomenclatur}

First, we define the missing energy for CC1p1h events with an outgoing muon and proton. The corresponding antineutrino variables are defined by exchanging the proton and neutron labels, $p \leftrightarrow n$, together with $\nu \leftrightarrow \bar{\nu}$. As shown in Ref.~\cite{Douqa2024}, the modified missing energy is related to the removal energy ($E_{rem}$) by
$\tilde{E}_m=E_{\rm rem}+T_{A-1}$,
where $T_{A-1}$ is the kinetic energy of the residual nucleus. Neglecting this recoil contribution (typically less than 3~MeV), we identify the quantity $S\equiv E_{\rm rem}\simeq\widetilde{E}_m$, and define
\begin{align}
S
=
E_\nu-E_\mu-E_p+M_n,
\qquad
p \leftrightarrow n,\quad
\nu \leftrightarrow \bar{\nu}.
\label{removal SRE}
\end{align}
 $E_\nu$ denotes the true event-by-event incoming-neutrino energy, $E_\mu$ and $E_p$ are the total energies of the outgoing muon and proton, respectively, and $M_n$ is the neutron mass. Here, ``event-level'' refers to the fact that $S$ is evaluated separately for each generated event using the true neutrino energy. Since such an event-by-event determination is not possible experimentally, in the following section we replace $S$ by an average value used as a constant in the reconstruction.
\subsection{Energy reconstruction}
\label{sec:reco_methods}

The standard energy reconstruction approximates the event-level quantity $S$ by its model average,
\begin{align}
S_{\rm RE}\equiv \bar S .
\end{align}
 Solving Eq.~\eqref{removal SRE} for $E_\nu$ then gives the reconstructed neutrino energy
\begin{align}
E_\nu^{\rm RE}
&=
S_{\rm RE}+E_\mu+E_p-M_n.
\label{EnuRE}
\end{align}
Equation~\eqref{EnuRE} defines the standard RE estimator used in this section. 
The values of $S_{\rm RE}$ reported for each generator setup in Table~\ref{tab:sre_mean} are truth-level benchmarks and are not directly measurable. Their numerical values depend on the nuclear model, the convention used to define the missing or removal energy, the kinematic level at which the outgoing-proton energy is evaluated, and, in phenomenological implementations, the experimental data to which the model is tuned.
In Sec.~\ref{sec:psiprimereco}, we introduce an alternative estimator, $S_{\rm RE}^{\rm eff}$, determined from the reconstructed superscaling distribution.

\subsection{Target-at-rest approximation}

We denote by $\mathrm{TAR}$ the approximation in which the struck initial-state nucleon is assumed to be at rest. At this stage, $E_\nu$ denotes the generator-true incoming-neutrino energy and is retained as an independent kinematic variable. The TAR hypothesis therefore constrains the outgoing-proton energy through $E_\nu$ and the outgoing-muon kinematics, but does not yet define a muon-only reconstructed quantity. Taking the beam direction as the longitudinal axis, the incoming-neutrino momentum is purely longitudinal, while $p_{T\mu}$ and $p_{z\mu}$ denote the transverse and longitudinal components of the outgoing-muon momentum. The momentum transfer is then
\begin{align}
\left|\vec q(E_\nu)\right|
=
\sqrt{p_{T\mu}^2+\left(E_\nu-p_{z\mu}\right)^2}.
\end{align}
Under the TAR hypothesis, the initial neutron is taken to be at rest,
\(\vec p_n^{\,\rm init}=0\). Momentum conservation at the interaction vertex then gives
\(\vec p_p^{\,\rm TAR}=\vec q\), where \(\vec q\) is the three-momentum transfer. The outgoing proton is taken to be on shell, so that its energy is
\[
E_p^{\rm TAR}
=
\sqrt{M_p^2+\left|\vec q(E_\nu)\right|^2},
\]
which is fully determined by the neutrino energy and the outgoing-muon kinematics. The resulting event-level removal-energy quantity is therefore
\begin{align}
S_{\rm LQE}
=
E_\nu-E_\mu-\sqrt{M_p^2+\left|\vec q(E_\nu)\right|^2}+M_n.
\label{SQE}
\end{align}

\begin{figure}[b]
    \centering
    \includegraphics[width=0.7\linewidth]{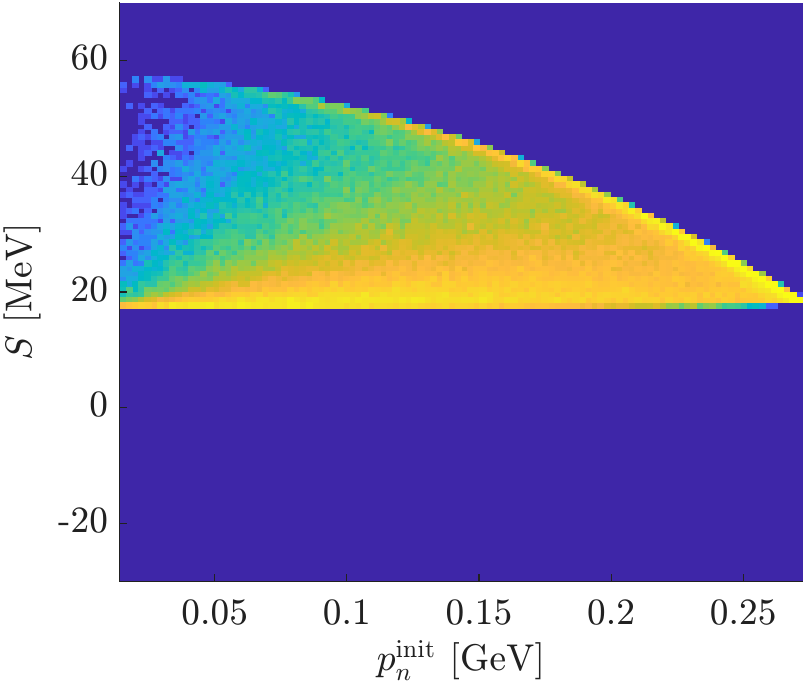}
    \includegraphics[width=0.7\linewidth]{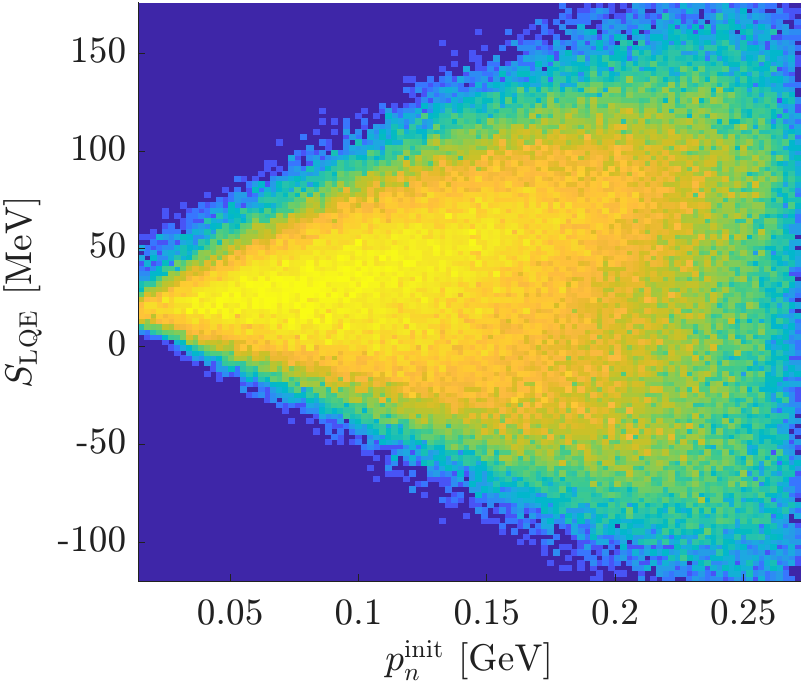}
  \caption{Event-level removal-energy variables as functions of the initial target-neutron momentum for CC1p1h events generated with NEUT using the LFG model. The top panel shows the vertex-level removal energy \(S\), while the bottom panel shows the corresponding quantity \(S_{\rm LQE}\) obtained under the target-at-rest approximation.}
    \label{fig:true removal}
\end{figure}

Figure~\ref{fig:true removal} illustrates the relation between the initial target-neutron momentum and the removal-energy variables for the LFG model. The vertex-level quantity \(S\) exhibits the characteristic LFG behavior, with a narrow distribution around the removal-energy scale and a sharp cutoff at \(p_n^{\rm init}\simeq k_F\), corresponding to the boundary of the carbon Fermi sphere. After imposing the target-at-rest approximation, the resulting quantity \(S_{\rm LQE}\) develops a cone-like structure: its distribution is narrow for small \(p_n^{\rm init} \approx 0 \), where the approximation remains accurate, and progressively broadens as the initial momentum approaches \(p_n^{\rm init} \approx k_F\).

The corresponding correlations for the considered nuclear models are shown in Appendix~\ref{App:Nuclearmodel}. The SF configurations retain the same qualitative behavior, but the sharp cutoff at \(k_F\) is replaced by a high-momentum tail extending beyond the nominal LFG Fermi-sphere boundary. These configurations consequently populate regions in which the target-at-rest reconstruction becomes less accurate, with implications for the extended superscaling tails discussed later. Differences among the configurations therefore reflect their different initial-nucleon momentum and removal-energy distributions.

\subsection{Lepton-only quasielastic reconstruction}
\label{fixingSQE}

In this case, only the outgoing-muon kinematics are assumed to be accessible. We refer to the resulting scheme as leptonic quasielastic (LQE) reconstruction. Quantities carrying the superscript or subscript LQE are obtained by imposing the quasielastic target-at-rest hypothesis and by replacing the event-level removal-energy quantity $S_{\rm LQE}$ with a fixed estimator $\bar S_{\rm LQE}$. The TAR hypothesis removes the dependence on an independently measured outgoing-nucleon energy, while fixing $S_{\rm LQE}=\bar S_{\rm LQE}$ closes the kinematics and allows the neutrino energy to be determined from the outgoing muon kinematics. The resulting LQE quantities therefore depend only on the outgoing-muon momentum and angle and on the constant $\bar S_{\rm LQE}$. For each event, the reconstructed neutrino energy $E_\nu^{\rm LQE}$ is then obtained as the physical solution of
\begin{align}
E_\nu^{\rm LQE}
&=E_\mu
+\sqrt{M_p^2+q(E_\nu^{\rm LQE})^2}
-M_n
+\bar S_{\rm LQE}.
\label{eq:SLQE_def}
\end{align}
Introducing the shorthand
\begin{align}
\widetilde M_n \equiv M_n-\bar S_{\rm LQE},
\end{align}
and isolating the square-root term in Eq.~\eqref{eq:SLQE_def}, we obtain
\begin{align}
E_{\nu}^{\rm LQE}
=
\frac{
m_{\mu}^{2}
-2E_{\mu}\widetilde M_n
+\widetilde M_n^{\,2}
-M_{p}^{2}
}{
2\left(E_{\mu}-\widetilde M_n-p_{\mu}\cos\theta_\mu\right)
}.
\label{eq:EnuLQE}
\end{align}
Under the target-at-rest approximation, the corresponding reconstructed proton energy is
\begin{align}
E_p^{\rm LQE}
&= \sqrt{M_p^2 + q(E_\nu^{\rm LQE})^2}
\notag\\
&= E_\nu^{\rm LQE}(\bar S_{\rm LQE}) - E_\mu + M_n - \bar S_{\rm LQE} .
\label{eq:EpLQE_closed}
\end{align}

For antineutrino interactions, the same reconstruction scheme applies after exchanging the neutron and proton labels, $n\leftrightarrow p$. In that case, the target-at-rest hypothesis yields the corresponding reconstructed outgoing-neutron energy $E_n^{\rm LQE}$ from the measured muon kinematics.

The dependence on \(\bar S_{\rm LQE}\) in Eq.~\eqref{eq:EnuLQE}, and therefore also in Eq.~\eqref{eq:EpLQE_closed}, is nonlinear. As a result, changing \(\bar S_{\rm LQE}\), or comparing it with \(S_{\rm RE}\), does not correspond to a simple constant shift of the reconstructed neutrino energy. For a fixed value of \(\bar S_{\rm LQE}\), there is a one-to-one mapping between the muon variables \((p_\mu,\theta_\mu)\) and the reconstructed variables \((E_\nu^{\rm LQE},E_p^{\rm LQE})\). The muon-kinematic distribution can therefore equivalently be studied in the bijectively related \((E_\nu^{\rm LQE},E_p^{\rm LQE})\) space.

In the literature, the parameter \(\bar S_{\rm LQE}\) is often called \(E_b\), for binding energy, and acts as an energy offset to the initial target-nucleon mass in the quasielastic reconstruction. We avoid the notation \(E_b\) to distinguish this effective reconstruction parameter from the nuclear binding energy. The peak positions extracted from the \(S_{\rm LQE}\) distributions are reported in Table~\ref{tab:sqe_peak}. Several reference values have been reported for the carbon energy-shift or removal-energy scale. Superscaling analyses of \((e,e')\) data typically obtain values around \(20\,\mathrm{MeV}\)~\cite{Donnelly1999}, while MINER\(\nu\)A quasielastic reconstruction analyses have used values closer to \(34\,\mathrm{MeV}\)~\cite{MINERvA:2013kdn}; the NEUT LFG implementation uses approximately \(25\)--\(27\,\mathrm{MeV}\) \cite{Bourguille2021}.
The model-averaged peak positions are approximately \(26\,\mathrm{MeV}\) for both neutrinos and antineutrinos; we adopt this value as a common reference unless otherwise stated. The precise choice of $\bar S_{\rm LQE}$ does not qualitatively affect the conclusions, since it is used only to define a common reconstruction baseline close to the peak scale of all models.

\subsection{Comparison of RE and LQE reconstructions}
\label{DEnureco}

We now combine the RE and LQE reconstruction schemes to define the neutrino-energy reconstruction bias. The true neutrino-energy reconstruction bias is defined as the difference between the true neutrino energy and the neutrino energy reconstructed under the LQE-TAR assumption,
\begin{align}
\Delta E_\nu
\equiv
E_\nu-E_\nu^{\rm LQE} .
\label{eq:DEnu_true}
\end{align}
Since the incoming neutrino energy is unknown, \(\Delta E_\nu\) cannot be measured directly but can be estimated using the RE reconstruction.
The RE estimator uses the outgoing-nucleon energy $E_p$, whereas $E_\nu^{\rm LQE}$ and $E_p^{\rm LQE}$ are inferred from the muon kinematics using the fixed parameter $\bar S_{\rm LQE}$. In the present comparison, $E_p$ and $S_{\rm RE}$ are both evaluated consistently at the interaction vertex before nuclear rescattering. Replacing the unknown true neutrino energy in Eq.~\eqref{eq:DEnu_true} by its RE estimator defines the reconstructed bias: 
\begin{align}
\Delta E_\nu^{\rm RE}
&\equiv
E_\nu^{\rm RE}
-
E_\nu^{\rm LQE}
\nonumber\\
&=
\bigl(E_p-E_p^{\rm LQE}\bigr)
+
\bigl(S_{\rm RE}-\bar S_{\rm LQE}\bigr)
\nonumber\\
&=
\Delta E_p
+
\bigl(S_{\rm RE}-\bar S_{\rm LQE}\bigr)
\label{eq:psiprime_num_reco}
\end{align}
where
\begin{align}
\Delta E_p
\equiv
E_p-E_p^{\rm LQE} .
\label{eq:DEp_definition}
\end{align}

Equation~\eqref{eq:psiprime_num_reco} establishes a direct link between the hadronic and muon-only reconstruction schemes. 
The reconstructed bias contains two contributions: the hadronic residual $\Delta E_p$, which compares the outgoing-nucleon energy with the LQE value inferred from the muon kinematics, and the sample-dependent constant reconstruction offset $S_{\rm RE}-\bar S_{\rm LQE}$. Through $E_p^{\rm LQE}$, $\Delta E_p$ retains a nonlinear dependence on $\bar S_{\rm LQE}$. The quantities $S_{\rm RE}$ and $\bar S_{\rm LQE}$ are distinct estimators of the removal-energy scale and are defined under different reconstruction assumptions. The standard estimator $S_{\rm RE}$ is obtained as the average of the event-level quantity $S$, whereas $\bar S_{\rm LQE}$ fixes the energy offset entering the muon-only target-at-rest reconstruction. Their difference is therefore not expected to vanish.

\begin{figure}[t]
    \centering
    \includegraphics[width=0.65\linewidth]{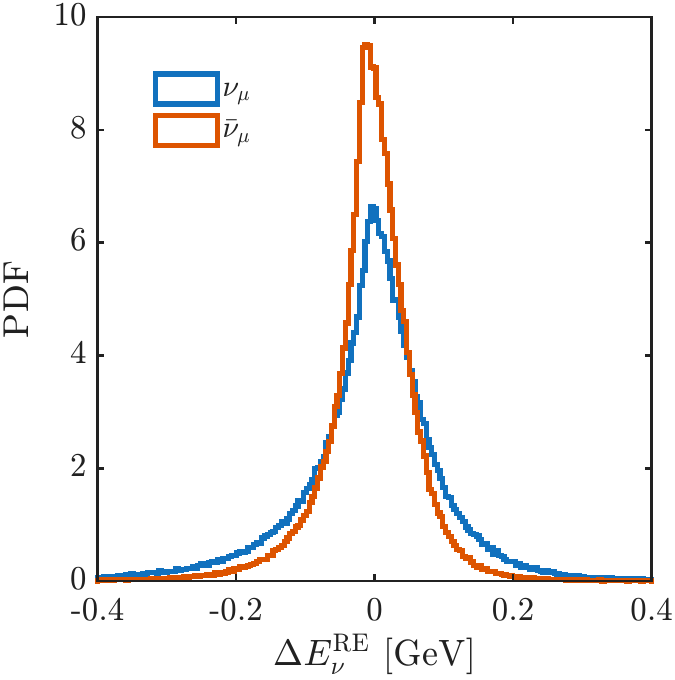}
    \includegraphics[width=0.65\linewidth]{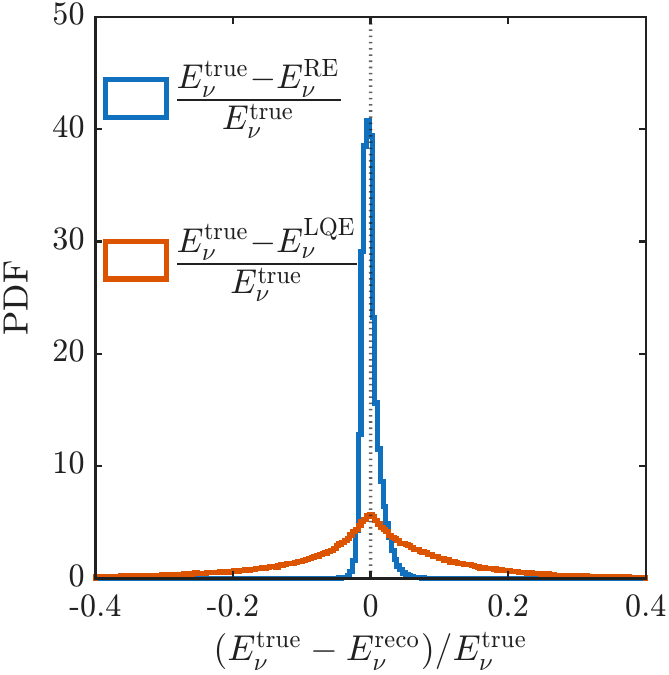}
 \caption{
Top: neutrino-energy reconstruction bias
$\Delta E_\nu^{\rm RE}
=
\Delta E_p+S_{\rm RE}-\bar S_{\rm LQE}$
for neutrino and antineutrino interactions generated with the NEUT LFG model.
Bottom: comparison of the corresponding relative RE and LQE neutrino-energy reconstruction residuals in the neutrino channel.}
    \label{fig:REQE_compare}
\end{figure}

Figure~\ref{fig:REQE_compare} shows the reconstructed neutrino-energy bias for neutrino and antineutrino interactions. The antineutrino reconstructed-bias distribution is also noticeably narrower than its neutrino counterpart; this difference will be interpreted in terms of the superscaling kinematics in Sec. \ref{subsubsec:L1_quasilinear}. For antineutrino interactions, the outgoing-neutron energy scale can be inferred under the target-at-rest assumption, but reconstructing \(\Delta E_\nu^{\rm RE}\) requires a measurement of the outgoing-neutron energy, which is experimentally challenging. The lower panel also shows that the hadronic RE estimator provides a narrower and more accurate reconstruction of the neutrino energy than the muon-only LQE estimator.

The observable \(\Delta E_p\) can be interpreted as a kinematic
energy-imbalance variable, comparing the outgoing-proton energy with
the proton energy inferred from the muon kinematics under the LQE
target-at-rest hypothesis. This construction is closely related to
inferred-proton and kinematic-imbalance observables previously studied
by T2K and in subsequent analyses of lepton--nucleon
correlations~\cite{PhysRevD.98.032003,PhysRevD.110.032019}. In the
present framework, however, the additional reconstruction offset
\(S_{\rm RE}-\bar S_{\rm LQE}\) establishes a direct connection between
this hadronic imbalance and the neutrino-energy reconstruction bias.
Since $\Delta E_p$ requires outgoing-nucleon information, it is only accessible in detector configurations with hadronic reconstruction. In the following, we express the superscaling variable $\psi'$ in terms of the reconstruction variables introduced above. This connection will allow us to derive new observables that depend only on the muon kinematics while retaining information about the neutrino-energy reconstruction bias.

\section{Superscaling Expansion and Muon-Only Observables}
\label{sec:L1_weak_EnuE}
We have presented two methods for reconstructing the neutrino energy from energy conservation. To connect these two reconstruction schemes to the neutrino-energy bias, we introduce the superscaling variable \(\psi'\). As established in Sec.~\ref{fixingSQE}, we use the common reference value \(\bar S_{\rm LQE}=26\,\mathrm{MeV}\). To ensure a common nuclear baseline across all configurations, we fix the carbon Fermi momentum to \(k_F=228\,\mathrm{MeV}\)~\cite{Maieron2002}. As in Sec.~\ref{sec:energy reco}, \(E_p\) denotes the hadronic-energy input at the chosen kinematic level; the relations derived below remain unchanged provided the reconstruction is defined consistently at that level.

\subsection{The Superscaling Variable $\psi'$ in neutrino-nucleus interactions}

When a lepton scatters from a many-body system in the impulse approximation, the inclusive nuclear response can be written, after an appropriate reduction, as the product of an elementary single-nucleon contribution and a nuclear scaling function \cite{Megias:2013qcw}
\begin{equation}
\frac{d^2\sigma_A}{d\Omega_\ell\,dE_\ell}
\;\simeq\;
\sigma_{\ell N}(\omega,|\vec q|)\; f(\psi')\,,
\end{equation}
where \(\sigma_{\ell N}\) denotes the effective single-nucleon contribution, including the relevant kinematic factors. A geometrical interpretation of the RFG scaling function is provided in Appendix~\ref{app:psi_RFG}.
Scaling of the first kind corresponds to the regime in which the
reduced response depends on \(\omega\) and \(|\vec q|\) only through
the single dimensionless variable \(\psi'(\omega,|\vec q|)\).
The superscaling variable \(\psi'\) was introduced by Donnelly and Sick~\cite{Donnelly1999,PhysRevLett.82.3212} as a refinement of the original RFG scaling variable \(\psi\) used by Alberico \textit{et al.}~\cite{PhysRevC.38.1801}.

We start from the generalized functional form of the superscaling variable for transitions between initial and final hadronic states of different masses~\cite{PhysRevC.69.035502},
\begin{align}
\psi'(E_\nu,E_\mu,\vec p_\mu)
&=
\chi_F\,
\frac{\lambda'-\tau'+\Delta M}{
\sqrt{
\left(1+\lambda'\rho\right)\tau'
+\kappa'\sqrt{\tau'\left(1+\tau'\rho^2\right)}
}}
\,,
\label{eq:psiprime_def}
\end{align}
and adapt its kinematic ingredients to charged-current neutrino scattering.
In contrast to electron scattering, charged-current quasielastic interactions involve a nucleon isospin transition \(n\to p\) (for neutrinos) or \(p\to n\) (for antineutrinos).
For notational simplicity, we use the neutrino transition \(n\to p\) as the reference configuration throughout this section. The corresponding antineutrino expressions are obtained by exchanging \(n\leftrightarrow p\). The difference between the initial- and final-state nucleon masses is encoded in the dimensionless correction
\begin{align}
\Delta M
&\equiv
\frac{M_n^2-M_p^2}{4M_n^2}.
\label{eq:rho_mass_correction}
\end{align}
For the charged-current \(n \to p\) transition, this correction is numerically small,
with \(\Delta M \simeq 6.9 \times 10^{-4}\). The corresponding correction for the
\(p \to n\) transition is obtained by interchanging the neutron and proton masses. We retain this mass-difference correction to keep the formalism as general as possible and to ensure a consistent extension to resonant scattering, where the corresponding mass shift is significantly larger, following the framework previously developed for electron scattering~\cite{Maieron2009}.
This correction is absent from the original Donnelly--Sick definition, which was developed for electromagnetic \(eA\) scattering and therefore involves no charged-current \(n\leftrightarrow p\) transition at the interaction vertex. We define:

\begin{align}
\kappa'
&=
\frac{|\vec q|}{2M_n},
&
\lambda'
&=
\frac{\omega-\bar{S}_{LQE}}{2M_n},
\nonumber\\
\tau'
&=
(\kappa')^2-(\lambda')^2,
&
\rho
&=
1-\frac{\Delta M}{\tau'},\\
\chi_F
&=
\frac{1}{\sqrt{\sqrt{1+\eta_F^2}-1}},
&
\eta_F
&=
\frac{k_F}{M_n}.
\label{eq:psi_kinematic_defs}
\end{align}
where $k_F$ is the target-dependent Fermi momentum, set to 228~MeV for all carbon samples following the prescription introduced in Ref.~\cite{Maieron2002}, and the mass term \(\Delta M\) encodes the charged-current
\(n\leftrightarrow p\) transition through \((M_n,M_p)\).
The energy and momentum transfers are expressed in terms of the incoming-neutrino energy and the outgoing-muon kinematics as
\begin{align}
\omega &= E_\nu - E_\mu,\qquad
|\vec q| = \sqrt{p_{T\mu}^2 + (E_\nu - p_{z\mu})^2},
\label{eq:qomega_muon}
\end{align}
with \(p_{T\mu}\) and \(p_{z\mu}\) the transverse and longitudinal components of the outgoing-muon momentum.
The superscaling variable $\psi'$ is designed such that its numerator vanishes for quasielastic target-at-rest events (CCQE-TAR), corresponding to a vanishing initial momentum of the struck nucleon, $p_N^{\rm init}=0$.

In practice, this condition is enforced through the cancellation of
\(\lambda'-\tau'+\Delta M\), which,
for a given muon kinematics, is equivalent to solving the CCQE-TAR kinematic equation for the incoming energy.
Denoting by \(E_\nu^{\rm LQE}(\bar{S}_{LQE})\) the muon-only energy estimator defined with an effective binding parameter \(\bar{S}_{LQE}\),
one has
\begin{align}
\lambda'\!\left(E_\nu^{\rm LQE};\bar S_{\rm LQE}\right)
-\tau'\!\left(E_\nu^{\rm LQE};\bar S_{\rm LQE}\right)
+\Delta M
=0.
\label{TAR eq}
\end{align}
Therefore, at fixed muon kinematics, the numerator of $\psi'$ reflects the reconstruction bias
$E_\nu - E_\nu^{\rm LQE}$ and vanishes when this bias is zero.
In an ideal CCQE 1p1h event with the target at rest, one has \(E_\nu=E_\nu^{\rm LQE}(\bar{S}_{LQE})\) and therefore the numerator vanishes. More physically, $\psi'$ can be interpreted as the relativistic counterpart of the $y$-scaling variable: quasielastic kinematics constrains the minimum longitudinal momentum of the initial nucleon along $\vec q$, with $k_F$ providing the characteristic momentum scale \cite{Amaro2021}. Thus, $\psi'$ maps the displacement from the quasielastic point onto the initial-nucleon momentum scale associated with Fermi motion, as discussed geometrically in Appendix~\ref{app:psi_RFG}.

\subsection{New parametrization of $\psi'$}
\label{sec:psireco}
As shown in the previous section, the $\psi'$ variable describes the phase-space distribution around the target-at-rest (TAR) point defined in Eq.~(\ref{TAR eq}). From the perspective of LQE reconstruction, the TAR condition corresponds to $E_\nu=E_\nu^{\text{LQE}}$, which motivates introducing the neutrino-energy reconstruction bias
$\Delta E_\nu=E_\nu-E_\nu^{\text{LQE}}$. The dependence of $\psi'$ on $E_\nu$ can therefore be replaced by a dependence on $\Delta E_\nu$, with $\psi'$ vanishing by construction when the reconstruction bias is zero:
\begin{align}
\psi'(E_\nu,\vec{p}_\mu)
&=
\psi'(\Delta E_\nu+E_\nu^{\text{LQE}},\vec{p}_\mu)
\equiv
\psi'(\Delta E_\nu,\vec{p}_\mu),
\\
&\psi'(\Delta E_\nu=0,\vec{p}_\mu)
=0
\qquad (\mathrm{TAR}).
\end{align}

Since \(\psi'\) can be expressed as a function of the neutrino-energy reconstruction bias, reconstructing \(\psi'\) requires reconstructing \(\Delta E_\nu\).

As derived in Sec.~\ref{DEnureco}, the true bias
\(\Delta E_\nu = E_\nu-E_\nu^{\rm LQE}\)
can be estimated through the RE reconstruction.

The RE-based reconstructible superscaling variable is then obtained by replacing the true neutrino-energy bias with its reconstructed estimator:
\begin{align}
\psi'(\Delta E_\nu,\vec p_\mu)
\longrightarrow
\psi'_{\rm reco}(\Delta E_\nu^{\rm RE},\vec p_\mu).
\label{eq:psiprime_num_reco2}
\end{align}

Equivalently, since
\(\Delta E_\nu^{\rm RE}
=
\Delta E_p+S_{\rm RE}-\bar S_{\rm LQE}\),
the reconstructed superscaling variable can be expressed in terms of the hadronic energy bias $\Delta E_p$ and the muon kinematics:
\begin{align}
\psi'_{\rm reco}
=
\psi'\!\left(
\Delta E_p+S_{\rm RE}-\bar S_{\rm LQE},
\vec p_\mu
\right).
\label{eq:psiprime_reco_DEp}
\end{align}
For antineutrino interactions, the equivalent relation is obtained through $n\leftrightarrow p$, with $\Delta E_p$ replaced by the outgoing-neutron residual $\Delta E_n$. Equation~(\ref{eq:psiprime_reco_DEp}) applies to the standard RE estimator introduced in Sec.~\ref{sec:reco_methods}.
\subsection{Taylor expansion of $\psi'$}
Once reconstructed, $\psi'$ becomes experimentally reconstructible, although its dependence on the outgoing-nucleon energy makes its measurement more challenging. This motivates a Taylor expansion around the TAR point $\Delta E_\nu =0$, or, after reconstruction, in the combination \(\Delta E_p+S_{\rm RE}-\bar S_{\rm LQE}\).
The expansion separates the experimentally challenging hadronic dependence from the coefficients $L_n$, which depend only on the muon kinematics and on the fixed nuclear inputs $M_n$, $M_p$, and $\bar S_{\rm LQE}$. By construction, the zeroth-order term vanishes, since the expansion is performed about the TAR point where $\psi'(\Delta E_\nu=0)=0$. Accordingly, $L_n$ has dimensions $[E^{-n}]$, so that each term in the expansion is dimensionless.
\begin{align}
\psi'(\Delta E_\nu,\vec p_\mu)
&=
\chi_F
\sum_{n=1}^{\infty}
L_n(\vec p_\mu)(\Delta E_\nu)^n ,
\\
\psi_{\rm reco}'(\Delta E_p,\vec p_\mu)
&=
\chi_F
\sum_{n=1}^{\infty}
L_n(\vec p_\mu)
\left[
\Delta E_p
+\bigl(S_{\rm RE}-\bar S_{\rm LQE}\bigr)
\right]^n .
\label{eq:taylor_psiprime}
\end{align}
The Taylor coefficients are defined by
\begin{align}
L_n(\vec p_\mu)
&\equiv
\left.
\frac{1}{n!}
\frac{\partial^n
\left[
\psi'(\Delta E_\nu,\vec p_\mu)/\chi_F
\right]}
{\partial(\Delta E_\nu)^n}
\right|_{\Delta E_\nu=0}.
\label{eq:Ln_def}
\end{align}
The first three analytical expressions for the $L_n$ observables used in this work are available in Appendix \ref{app:Ln_analytical}.

\begin{figure}[b]
    \centering
    \includegraphics[width=0.75\linewidth]{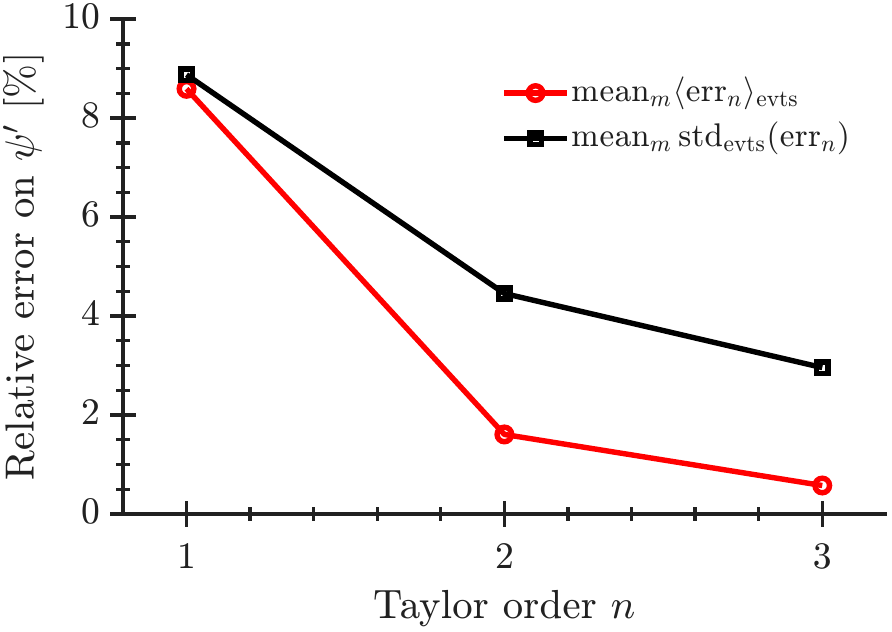}
    \caption{Taylor-series convergence for $\psi'$: model average of the
event-level mean normalized residual and model average of the
corresponding event-level standard deviation, shown as a function
of the truncation order $n$.}
    \label{fig:conv}
\end{figure}

\subsubsection{Convergence of the Taylor expansion}
The Taylor expansion of $\psi'$ around the TAR point converges well over the relevant phase space.
Figure~\ref{fig:conv} summarizes the convergence through the normalized residual.
\begin{align}
\epsilon_n \equiv
100\,\frac{\bigl|\psi'_{\rm true}-\psi'^{(n)}_{\rm Taylor}\bigr|}
{\bigl|\psi'_{\rm true}\bigr|+\epsilon_0}\,,
\end{align}
where a fixed numerical offset \(\epsilon_0 = 10^{-6}\) is introduced to regularize the denominator for events with \(\psi'_{\rm true}\simeq 0\).
This value is sufficiently small to have a negligible impact
away from the immediate vicinity of the TAR point. The figure shows the average over the generator configurations introduced in Sec.~\ref{Sec:NuclearModels} of the event-level mean of \(\epsilon_n\), together with the corresponding standard deviation, as a function
of the truncation order \(n\). Consistent with the behavior expected from a local Taylor approximation of a smooth function, the normalized residual $\epsilon_n$ decreases rapidly as the truncation order $n$ increases. At first order, the model-averaged residual is already below $10\%$, and it falls below $1\%$ once terms up to third order are included.

\subsubsection{Local expansion of $\psi'$ and the role of $L_1$, $L_2$, and $L_3$}
\label{L1expl}

In the quasielastic (QE) regime, $\psi'$ can be expanded around the target-at-rest (TAR) point, $\Delta E_\nu \equiv E_\nu^{\mathrm{true}}-E_\nu^{\mathrm{LQE}} = 0$, which corresponds to the local zero of the superscaling variable. To third order, one obtains
\begin{align}
\psi^{\prime}_{\rm true}
&=
\chi_F\left[
L_1\,\Delta E_\nu
+L_2\,(\Delta E_\nu)^2
+L_3\,(\Delta E_\nu)^3
\right]
+\mathcal{O}(\Delta E_\nu^4).
\label{DeltaE}
\end{align}

\begin{figure}[t]
    \centering
    \includegraphics[width=0.9\linewidth]{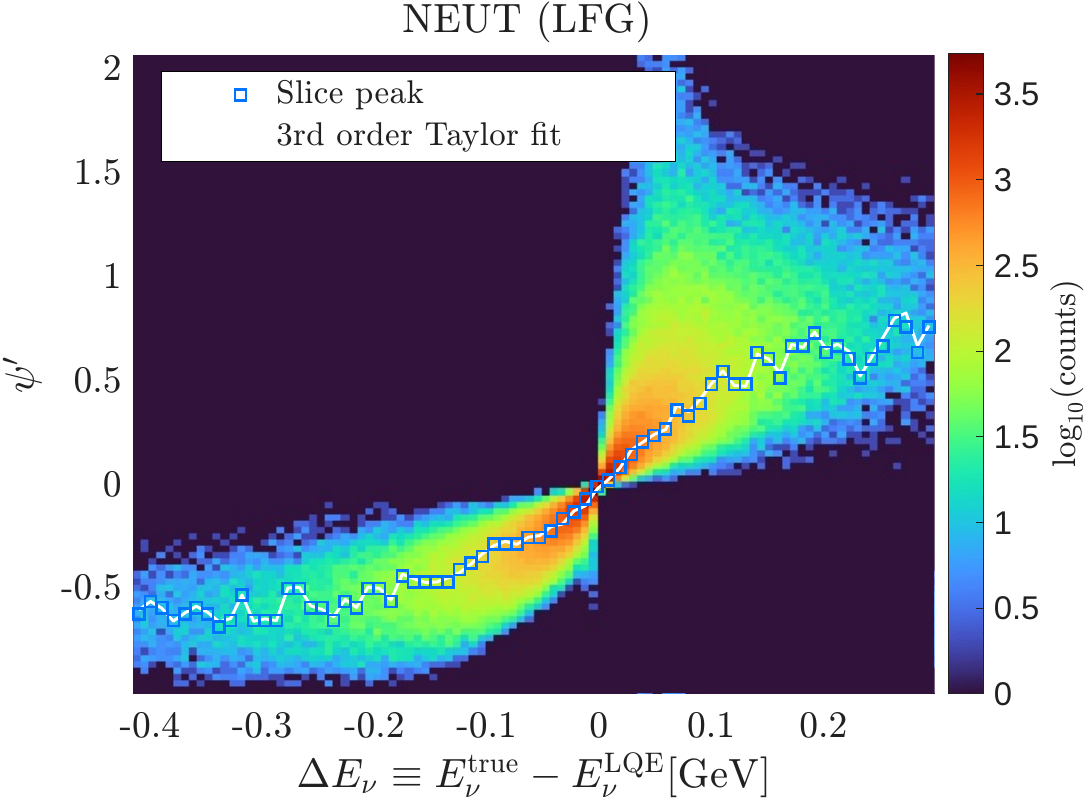}
\caption{
True superscaling variable $\psi'_{\rm true}$ as a
function of the neutrino-energy reconstruction bias
$\Delta E_\nu$ for the NEUT LFG model. The color map
shows the full two-dimensional distribution of
$\psi'_{\rm true}$. The blue squares indicate the peak positions obtained from Gaussian fits to the $\psi'$ distribution in each $\Delta E_\nu$ slice, while the solid curve shows the corresponding fitted peak positions for the full third-order Taylor approximation $\psi_{\rm Taylor}^{\prime(3)}$. The complete
event-by-event convergence of the Taylor expansion is quantified
through the normalized residual in Fig.~\ref{fig:conv}.
} \label{psiprimefit}
\end{figure}

The interpretation of the three coefficients is illustrated by Fig.~\ref{psiprimefit}. In each slice of $\Delta E_\nu$, the peak position of the $\psi'$ distribution is extracted using an iterative Gaussian least-squares fit, with the fit range progressively narrowed around the peak until the fitted peak position converges.
The coefficient $L_1$ controls the local linear slope of $\psi'$ around the TAR point $\Delta E_\nu=0$. 

In this near-QE region, the dominant behavior is $\psi' \simeq \chi_F\,L_1\,\Delta E_\nu$,
so that $L_1$ determines how rapidly $\psi'$ moves away from zero when a small neutrino-energy bias is introduced. 
This interpretation is consistent with Fig.~\ref{psiprimefit}. Within approximately \(\pm100~\mathrm{MeV}\) of the TAR point, the distribution is nearly linear and dominated by \(L_1\).

The coefficient $L_2$ controls the first departure from this linear behavior. 
As $|\Delta E_\nu|$ becomes non-negligible, the quadratic term generates the first curvature of the ridge and therefore controls the local deviation from linearity.
In this sense, $L_2$ quantifies how rapidly the linear approximation breaks down as one moves away from the TAR region.

The quadratic term provides the leading contribution that breaks the local antisymmetry between positive and negative reconstruction biases.
The cubic term is odd in $\Delta E_\nu$ and provides the leading correction to the asymmetric shape of $\psi'$ beyond the quadratic approximation.
As a result, $L_3$ is expected to be most relevant in the more extreme regions of phase space, far from the TAR condition, where the cubic term becomes comparable to or larger than the quadratic correction.

\subsection{The $L_1$ observable}
The leading Taylor coefficient depends only on the outgoing-muon kinematics and is given by 
\begin{equation}
\resizebox{1\linewidth}{!}{$
L_1(\vec p_\mu)
=
\frac{\sqrt{2}}{M_n}
\frac{
\left(\widetilde{M}_n+p_{z\mu}-E_\mu\right)^2
}{
\sqrt{
\left[
p_{T\mu}^2+
\left(M_p-\widetilde{M}_n-p_{z\mu}+E_\mu\right)^2
\right]
\left[
p_{T\mu}^2+
\left(M_p+\widetilde{M}_n+p_{z\mu}-E_\mu\right)^2
\right]
}
}
$}
\label{eq:L1_muon}
\end{equation}
where $\widetilde{M}_n = M_n - \bar S_{\rm LQE}$.
Its asymptotic behavior is characterized by a large but finite enhancement for forward muons with vanishing transverse momentum, while it approaches zero in the opposite kinematic limit.
\begin{align}
&\lim_{\substack{p_{T\mu}\to0\\E_\mu-p_{z\mu}\to0}}
L_1(\vec p_\mu)
=
\frac{\sqrt{2}}{M_n}
\frac{\widetilde{M}_n^{\,2}}
{\left|M_p^2-\widetilde{M}_n^{\,2}\right|}, \\
&
\lim_{E_\mu-p_{z\mu}\to \widetilde{M}_n}
L_1(\vec p_\mu)
=0.
\label{eq:L1_limits}
\end{align}
Using the TAR relation
$E_p^{\rm LQE}=E_\nu^{\rm LQE}-E_\mu+\widetilde{M}_n$,
Eq.~\eqref{eq:L1_muon} can be rewritten as
\begin{align}
L_1
=
\frac{
\widetilde{M}_n-(E_\mu-p_{z\mu})
}{
\sqrt{2}M_n
\sqrt{\left(E_p^{\rm LQE}\right)^2-M_p^2}
}.
\end{align}
Since
$
E_\mu-p_{z\mu}
=
\frac{Q_{\rm LQE}^2+m_\mu^2}
{2E_\nu^{\rm LQE}},
$
the high-energy limit
$(Q_{\rm LQE}^2+m_\mu^2)/(2E_\nu^{\rm LQE})\ll\widetilde{M}_n$,
together with $\bar S_{\rm LQE}\ll M_n$, gives
\begin{align}
L_1
\simeq
\frac{1}
{\sqrt{2}\sqrt{\left(E_p^{\rm LQE}\right)^2-M_p^2}}.
\label{eq:L1_asymptotic_pp}
\end{align}
Therefore, for $Q_{\rm LQE}^2\gg M_p^2+\widetilde{M}_n^2$,
\begin{align}
L_1
\simeq
\frac{\sqrt{2}M_n}{Q_{\rm LQE}^2}.
\label{eq:L1_asymptotic_Q2}
\end{align}

This asymptotic behavior explains the enhancement of $L_1$ toward low
$Q^2$ and its association with forward, low-$p_{T\mu}$ kinematics.

\begin{figure}[t]
    \centering
    \includegraphics[width=0.7\linewidth]{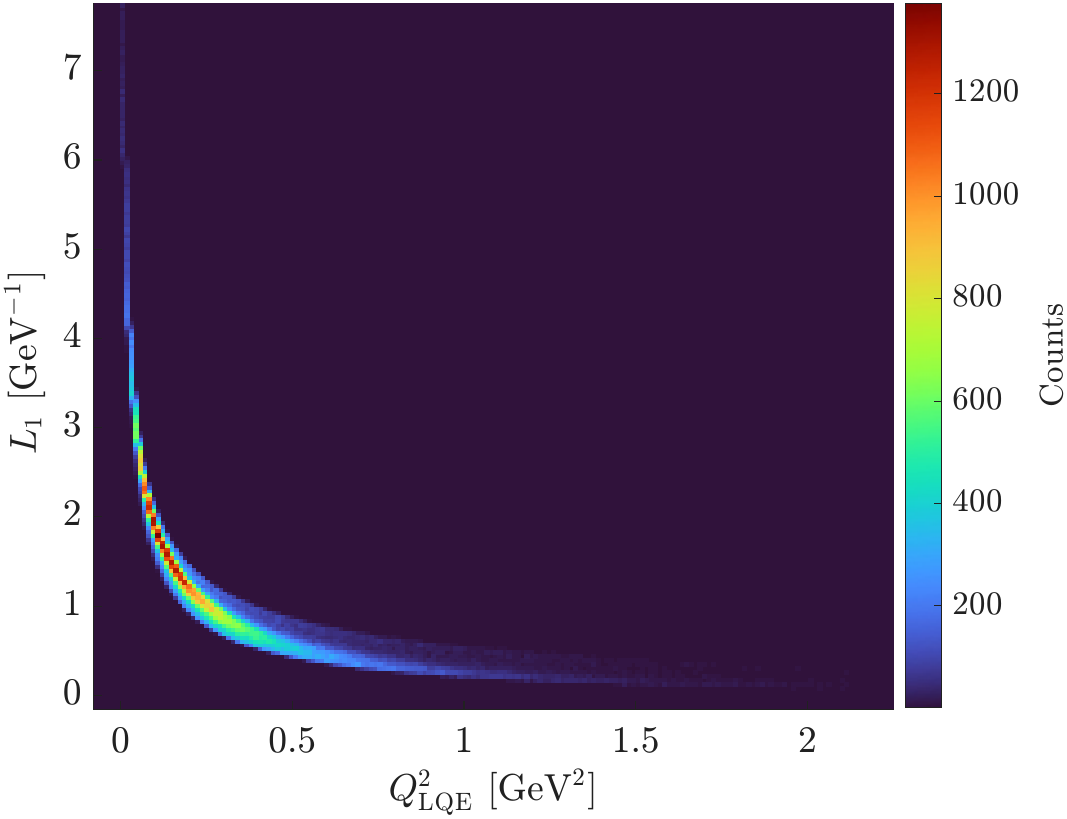}
    \caption{
Correlation between the first-order coefficient \(L_1\) and the
reconstructed four-momentum transfer \(Q_{\rm LQE}^2\) for
true CC1p1h events generated with the NEUT LFG model.
}
    \label{fig:L1Q}
\end{figure}

Figure~\ref{fig:L1Q} confirms the expected inverse correlation between $L_1$ and $Q_{\rm LQE}^2$. As the reconstructed momentum transfer increases, the reconstructed outgoing nucleon momentum also increases, leading to the asymptotic scaling $L_1 \propto 1/Q_{\rm LQE}^2$.
Although \(L_1\) exhibits an asymptotic correlation with \(Q_{\rm LQE}^2\), the two quantities are not equivalent event by event and have different distributions. More importantly, \(L_1\) follows directly from the local expansion of the superscaling variable and, as shown in the next section, defines a target-dependent bound on \(\Delta E_\nu\) through the superscaling factor \(\chi_F\), whereas \(Q_{\rm LQE}^2\) does not by itself yield the corresponding
superscaling bound.

\subsection{Constraining the neutrino-energy reconstruction bias with \(L_1\) and \(L_2\)}
\label{sec:L1DEnu}
\begin{figure}[t]
    \centering
    \includegraphics[width=0.7\linewidth]{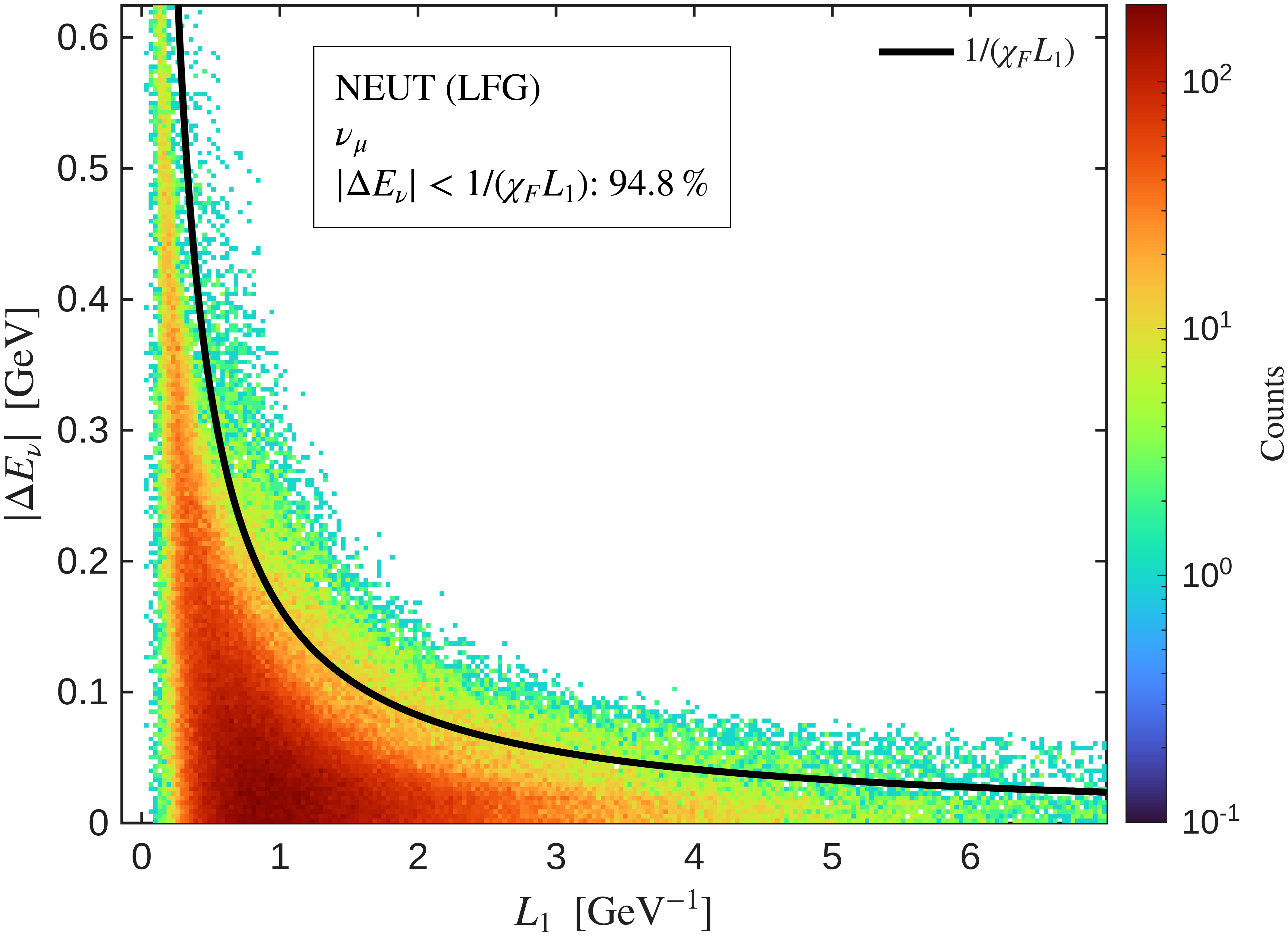}
    \includegraphics[width=0.65\linewidth]{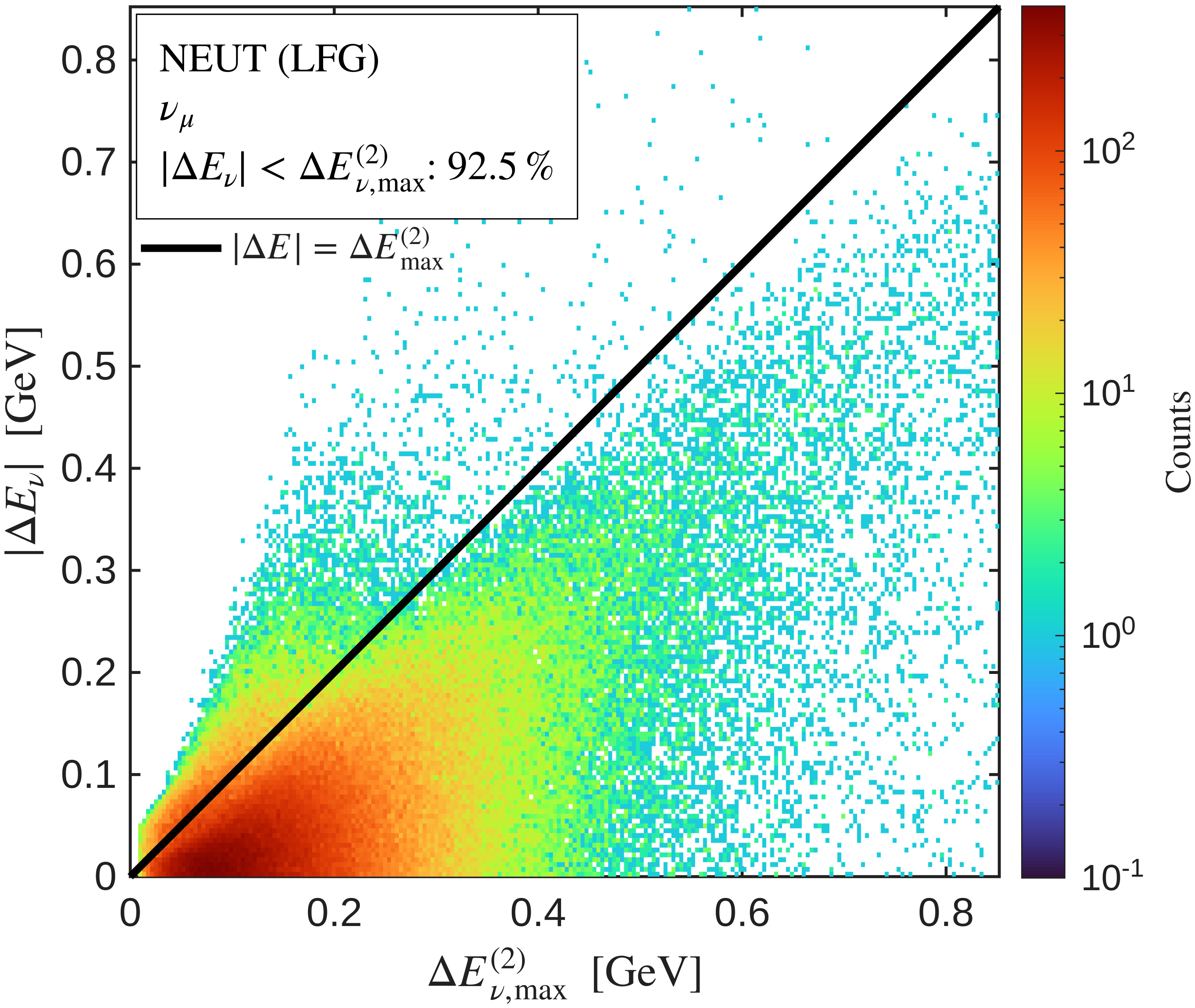}
\caption{Density distributions illustrating the superscaling constraints on the neutrino-energy reconstruction bias. \textbf{Top:} distribution in the $(L_1,|\Delta E_\nu|)$ plane with the first-order bound $|\Delta E_\nu|=1/(\chi_F L_1)$. \textbf{Bottom:} comparison between the true neutrino-energy bias and the corresponding second-order bound. The inclusion of the quadratic term yields a more restrictive event-by-event constraint on the allowed reconstruction bias. Both panels use the NEUT LFG $\nu_\mu$ sample.}
    \label{fig:L1DEnu}
\end{figure}

When the hadronic final state is not measured, the event-by-event bias of the LQE neutrino-energy reconstruction cannot be determined directly. Within the superscaling framework, however, the quasielastic scaling region associated with \(|\psi'|<1\) provides a kinematic constraint on this bias~\cite{Donnelly1999}. Using the first-order expansion of \(\psi'\), this condition gives
\begin{align}
|\psi'|
&\simeq
\chi_F L_1|\Delta E_\nu|<1,
\nonumber\\
|\Delta E_\nu|
&<
\Delta E_{\nu,\max}^{(1)}
\equiv
\frac{1}{\chi_F L_1}.
\label{eq:DeltaEnu_bound_L1}
\end{align}
Thus, $L_1$ provides an event-by-event upper bound on the neutrino-energy reconstruction bias and indicates whether $E_\nu^{\rm LQE}$ lies in a kinematic region where it is expected to be a reliable proxy for the true neutrino energy.
Figure~\ref{fig:L1DEnu} provides an event-by-event Monte Carlo validation of the analytical superscaling bounds. In the top panel, the simulated events follow the expected first-order envelope in the \((L_1,|\Delta E_\nu|)\) plane, with more than \(90\%\) lying below the boundary \(1/(\chi_F L_1)\). Events outside the analytical envelope correspond predominantly to the extended tails beyond the nominal superscaling region \(|\psi'|<1\). The constraint can be refined by including the second-order term:
\begin{align}
|\psi'|
&\simeq
\chi_F
\left|
L_1\Delta E_\nu
+
L_2(\Delta E_\nu)^2
\right|
<1.
\label{eq:DeltaEnu_bound_L2_condition}
\end{align}
From the analytical expressions given in
Appendix~\ref{app:Ln_analytical}, and restricting to the physical kinematic domain, \(L_1\) is strictly positive and \(L_2\) is strictly negative. The most restrictive symmetric solution of
Eq.~(\ref{eq:DeltaEnu_bound_L2_condition}) therefore gives
\begin{align}
|\Delta E_\nu|
&<
\Delta E_{\nu,\max}^{(2)}       
\equiv
\frac{
\sqrt{L_1^2+4|L_2|/\chi_F}-L_1
}{
2|L_2|
}.
\label{eq:DeltaEnu_bound_L2}
\end{align}
The bottom panel in Figure~\ref{fig:L1DEnu} confirms that the second-order expression provides an event-by-event bound in the simulation. 
At a near detector, this constraint can be tested through the correlation between $L_1$ and the reconstructed bias
$\Delta E_\nu^{\rm RE}$,
since all these quantities can be reconstructed from the outgoing-muon and proton kinematics.
\section{Nuclear model sensitivity of $\psi'$ and $L_n$ muon-only observables}
\label{sec:model_comparison}
\subsection{Nuclear models}
\label{Sec:NuclearModels}
To systematically evaluate how different nuclear models affect neutrino-interaction predictions, we carry out a comparative analysis within the NUISANCE framework~\cite{Stowell:2016jfr}, using ND280-like $\nu_\mu$ and $\bar{\nu}_\mu$ CC1p1h samples on a carbon target, corresponding to $n \to p$ and $p \to n$ transitions, respectively. Both channels are referred to collectively as CC1p1h throughout this work. Further details of the generator configurations and flux comparisons are given in Ref.~\cite{Dolan}.

The samples are generated with several nuclear-model implementations \cite{Wret2024NuIntPracticum}: \textsc{GENIE v3}~\cite{Andreopoulos:2009rq,PhysRevD.104.072009}, using the AR23 (LFG) \cite{PhysRevC.83.045501}, CRPA21 (HF-CRPA) \cite{Dolan2022}, and G21 (SuSAv2) \cite{Dolan2020} configurations; \textsc{NEUT}~6.1.4~\cite{Hayato2021}, using LFG \cite{Bourguille2021,PhysRevD.88.113007}, SF \cite{PhysRevD.72.053005}, and energy-dependent relativistic mean field (ED-RMF) \cite{f7x5-snmz}; and \textsc{NuWro}~\cite{Juszczak:2005zs}, using both LFG and SF nuclear initial-state models.

Nuclear rescattering (NrS), referring here to the intranuclear-cascade component of final-state interactions, modifies the outgoing nucleon as it propagates through the nuclear medium, affecting both its kinematics and the final-state topology. The vertex-level reconstruction quantities introduced in Sec.~\ref{sec:energy reco} are first studied before NrS. NrS is modeled through the intranuclear-cascade prescriptions implemented in each event generator~\cite{Salcedo:1987md,Yan:2026dyp,Liu:2025hpl,Golan:2012rfa,Dytman:2011zz}. The reconstructed superscaling observables considered later are then evaluated after NrS using the final-state muon and nucleon kinematics.

For the post-NrS study in the neutrino channel, we retain generator-level CC1p1h events containing exactly one final-state proton after nuclear rescattering, and use its total energy as the
hadronic input to the reconstruction. This generator-level prescription is used only to introduce the kinematic effects of nuclear rescattering into the hadronic reconstruction and should not be interpreted as an experimental nucleon-energy reconstruction. A realistic measurement would additionally depend on the observable final-state topology, detector acceptance and resolution, and the experimental hadronic-energy reconstruction strategy.
Such effects complicate the reconstruction of observables that depend on the outgoing-nucleon kinematics and provide additional motivation for deriving muon-only observables from the superscaling framework. 
\subsection{The truth-level \(\psi'\)}
\label{subsec:baseline_averages}

As discussed in Sec.~\ref{fixingSQE}, we fix the reference value \(\bar S_{\rm LQE}=26\)~MeV, corresponding to the model-averaged peak estimate across the generator setups considered.
Individual generators, however, favor values differing by approximately \(\pm 15\)~MeV from this reference. For example, NuWro SF gives \(\bar S_{\rm LQE}^{\nu}=42\)~MeV and \(\bar S_{\rm LQE}^{\bar\nu}=36\)~MeV, whereas GENIE v3 SuSAv2 gives \(16\)~MeV for both channels (see Table~\ref{tab:sqe_peak}).
This choice induces a truth-level misalignment of the \(\psi'\) peak positions across models, since we do not tune \(\bar S_{\rm LQE}\) as in the original superscaling analyses of electron scattering, where the corresponding parameter \(E_{\rm shift}\) is adjusted to center the TAR quasielastic peak at \(\psi'=0\).

We deliberately do not tune $\bar S_{LQE}$ here for two reasons. 
First, in neutrino experiments \(\psi'\) cannot be constructed in the same way as in \((e,e')\) scattering because the incoming lepton energy is unknown. Instead, we treat $\bar S_{\rm LQE}$ as a fixed external input and tune the effective reconstruction parameter $S^{\rm eff}_{\rm RE}$ such that the quasielastic target-at-rest (TAR) peak of the reconstructed $\psi'$ distribution is centered at zero.

\begin{figure}[t]
    \centering
    \includegraphics[width=0.75\linewidth]{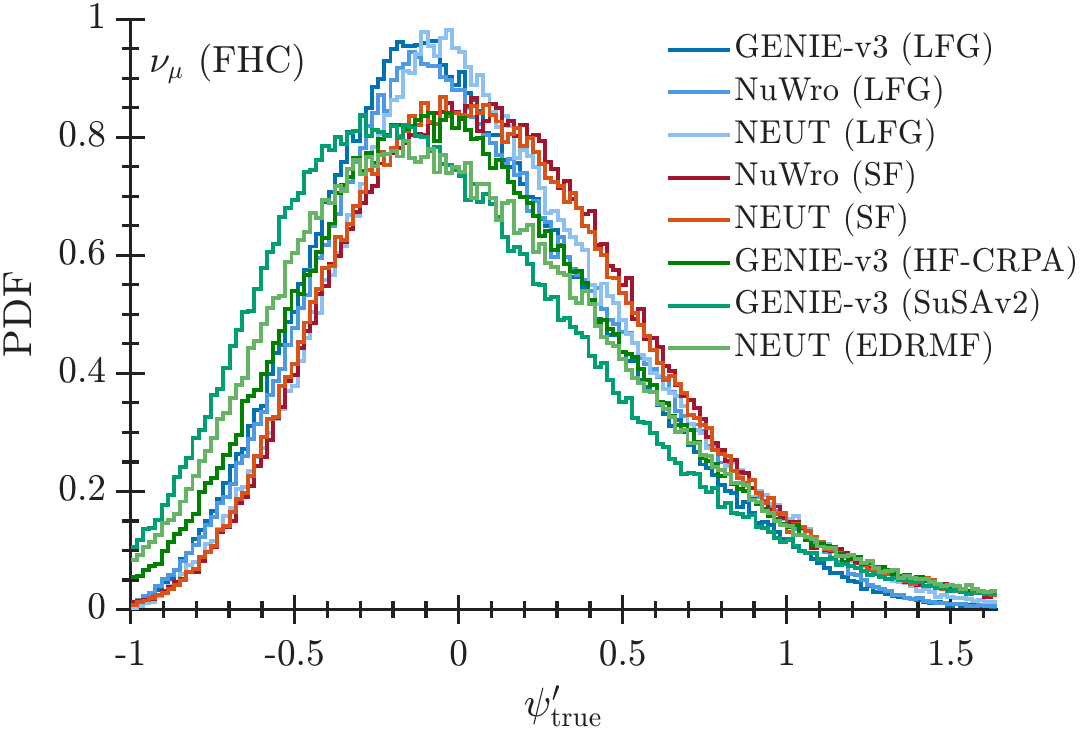}
       \includegraphics[width=0.75\linewidth]{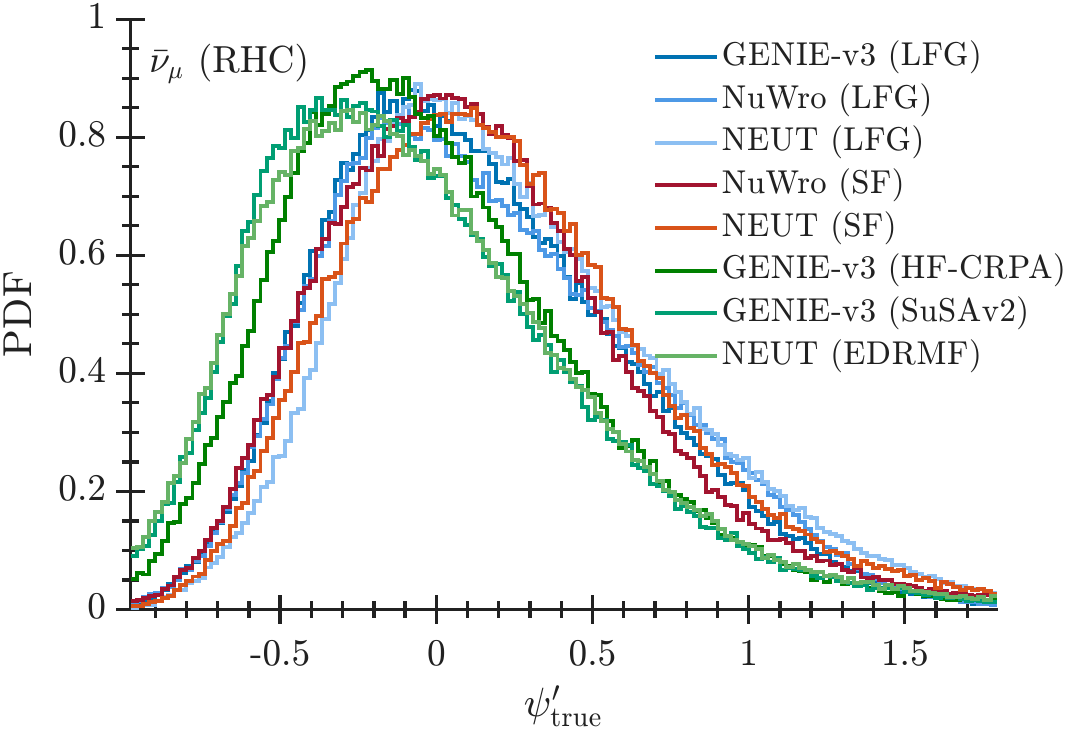}
    \caption{Truth-level superscaling variable \(\psi'\) for \(\nu_\mu\) (top) and \(\bar{\nu}_\mu\) (bottom), computed for all generator setups using the common reference \(\bar S_{\rm LQE}=26\)~MeV. Since the models yield different peak positions of \( S_{\rm LQE}\) (Table~\ref{tab:sqe_peak}), this common choice produces a visible model-dependent displacement of the \(\psi'\) peaks away from zero.}
    \label{psiprimetrue}
\end{figure}

This provides the practical handle required to realign the reconstructed $\psi'$ distributions. Second, varying $\bar S_{\rm LQE}$ modifies the muon-based reconstruction $E_\nu^{\rm LQE}$ nonlinearly, whereas $S_{\rm RE}^{\rm eff}$ enters the reconstructed energy linearly and absorbs the average model-dependent shift affecting the hadronic reconstruction, including the effects of nuclear rescattering. Figure~\ref{psiprimetrue} illustrates the resulting model-dependent displacement of the truth-level $\psi'$ peaks when the common reference value $\bar S_{\rm LQE}=26$~MeV is used. In the next subsection, this displacement is absorbed into $S_{\rm RE}^{\rm eff}$, bringing the reconstructed $\psi'$ distributions into a common reference frame. The $\nu_\mu$ and $\bar{\nu}_\mu$ samples are generated using the forward-horn-current (FHC) and reverse-horn-current (RHC) fluxes, respectively.
\label{subsec:baseline_averages}

\subsection{The reconstructed \(\psi'\)}
\label{sec:psiprimereco}

As shown in Sec.~\ref{sec:psireco}, $\psi'$ can be reconstructed by replacing the neutrino energy bias $\Delta E_\nu$ with the hadronic energy bias $\Delta E_p+S_{\rm RE}-\bar S_{\rm LQE}$ (Eqs.~\eqref{eq:psiprime_num_reco} and~\eqref{eq:psiprime_num_reco2}).  The corresponding antineutrino expression is obtained by interchanging the neutron and proton quantities, $n\leftrightarrow p$. When the outgoing-nucleon kinematics are evaluated after nuclear rescattering, the standard vertex-level estimator $S_{\rm RE}=\overline S$ need not provide the appropriate reconstruction offset. We therefore determine an alternative estimator, $S_{\rm RE}^{\rm eff}$, directly from the reconstructed superscaling distribution by requiring the fitted $\psi'$ peak to be centered at zero.
For each generator setup, we scan the calibration parameter $S^{\rm eff}_{\rm RE}$, which is not interpreted as a physical separation energy, over a fixed interval with a constant step, recompute the reconstructed $\psi'$ distribution, and extract its peak position using an iterative Gaussian fit.

The optimal value $S_{\rm RE}^{\rm eff}$ is then defined as the value that yields a vanishing fitted peak.
The resulting tuned values are reported in Table~\ref{tab:sre_tuned_psiprime}.

\begin{table}[b]
\centering
\caption{Tuned \(S^{\rm eff}_{\rm RE}\) values obtained by centering the reconstructed \(\psi'\) peak at zero for samples including NrS.}
\label{tab:sre_tuned_psiprime}
\begin{tabular}{lcc}
\hline
Model & $S_{\rm RE}^{\rm eff}(\nu)$ [MeV] & $S_{\rm RE}^{\rm eff}(\bar\nu)$ [MeV] \\
\hline
GENIE (LFG)      & $38$ & $26$ \\
NuWro (LFG)      & $33$ & $26$ \\
NEUT (LFG)       & $33$ & $26$ \\
NuWro (SF)       & $38$ & $26$ \\
NEUT (SF)        & $38$ & $26$ \\
GENIE (HF-CRPA)  & $12$ & $10$ \\
GENIE (SuSAv2)   & $37$ & $20$ \\
NEUT (ED-RMF)    & $52$ & $42$ \\
\hline
Average          & $35$ & $25$ \\
\hline
\end{tabular}
\end{table}
 These values are distinct from the standard vertex-level estimator $S_{\rm RE}=\overline{S}$ reported in Appendix~\ref{App:Nuclearmodel}, since they are obtained from post-NrS kinematics using a different calibration criterion.

\begin{figure}[t]
    \centering
    \includegraphics[width=0.75\linewidth]{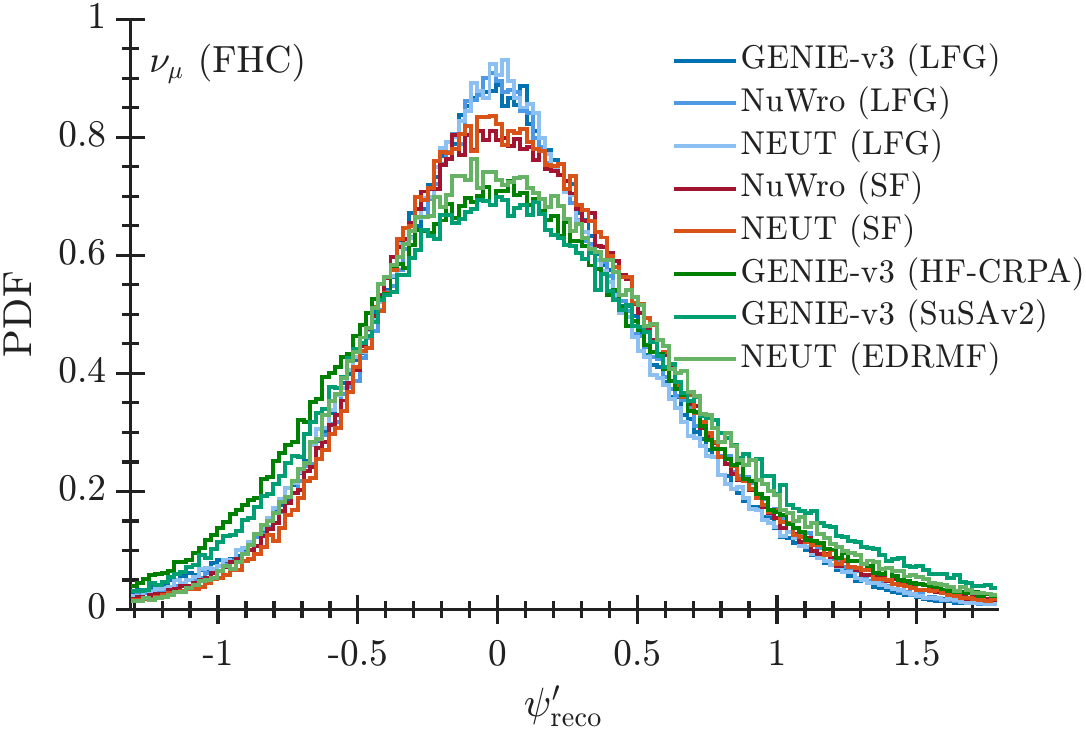}
    \includegraphics[width=0.75\linewidth]{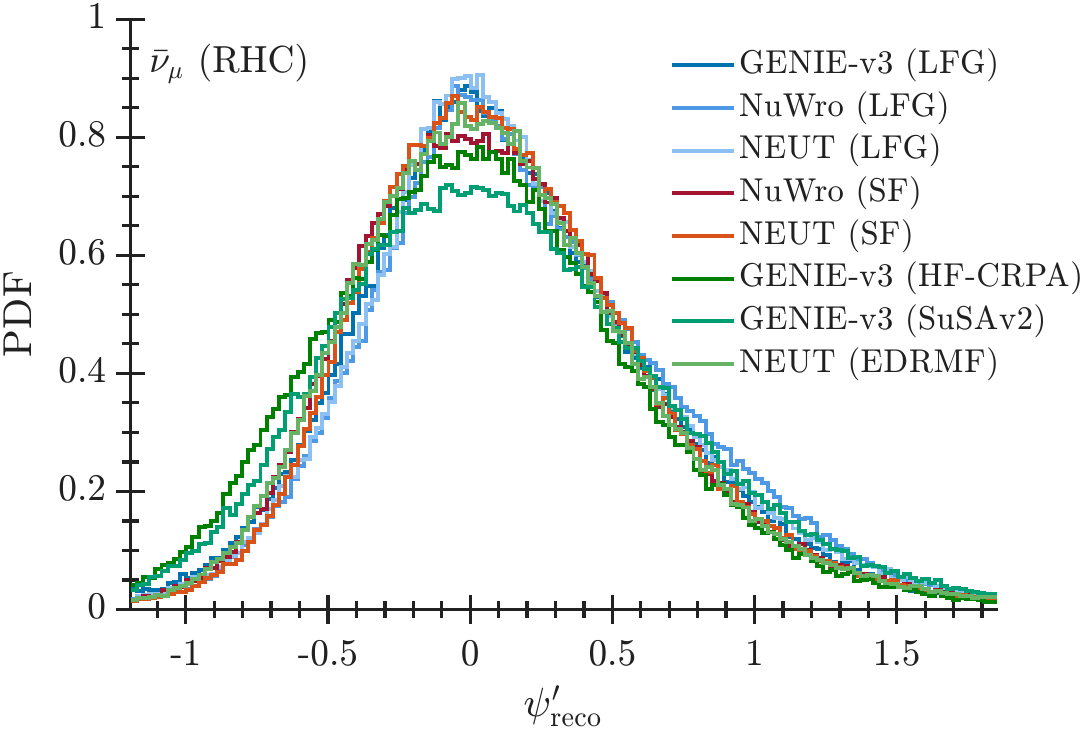}
    \caption{Reconstructed superscaling variable \(\psi'\) after tuning \(S^{\rm eff}_{\rm RE}\) to center the distribution at \(\psi'=0\), shown as PDFs for all generator setups for \(\nu_\mu\) (top) and \(\bar{\nu}_\mu\) (bottom). The antineutrino distribution is reconstructed at the Monte Carlo level using the outgoing-neutron energy.}
    \label{fig:psiprime_reco_overlays}
\end{figure}

The tuned reconstructed superscaling variable, which relies on the outgoing nucleon energy and is therefore directly accessible mainly in the neutrino channel, already exhibits visible model-dependent shape differences, as shown in Fig.~\ref{fig:psiprime_reco_overlays}. In practice, this observable is much harder to access for antineutrino interactions, since it would require a reliable reconstruction of the outgoing neutron energy, which is generally not available with sufficient precision in current detectors.

Figure~\ref{fig:psiprime_reco_overlays} shows that the tuned \(\psi'\) distributions cluster primarily according to the underlying nuclear-model family rather than the generator implementation. Because $S_{\rm RE}^{\rm eff}$ is tuned separately for each setup, these differences correspond to residual shape variations after the peak-position offset has been removed. The LFG predictions from \textsc{GENIE}, \textsc{NuWro}, and \textsc{NEUT} largely overlap, as do the two SF predictions. The HF-CRPA, SuSAv2, and ED-RMF distributions form a third group despite their different implementations and effective binding-energy prescriptions.
Once the overall peak displacement is absorbed into \(S^{\rm eff}_{\rm RE}\), the residual shape of the reconstructed \(\psi'\) distribution remains sensitive to the nuclear dynamics encoded by each model family. The LFG, SF, and mean-field-based descriptions remain visibly separated, while differences between generators implementing similar nuclear models are comparatively reduced.
The separation between these model families is correlated with
differences in their initial-state momentum and energy-balance prescriptions. In the LFG models, the sharp restriction \(p_n^{\rm init}\lesssim k_F\) confines most events to the nominal superscaling region. The SF configurations instead contain extended high-momentum components, which contribute directly to the longer \(\psi'\) tails. For HF-CRPA and SuSAv2, the generator-level
treatment of the initial-nucleon energy and event-by-event energy
balance also affects the mapping into \(\psi'\), as discussed in
Appendix~\ref{App:Nuclearmodel}.

Although the reconstructed antineutrino $\psi'$ is not experimentally accessible in the same way, its behavior in simulation provides a useful cross-check of the model interpretation. In the antineutrino channel, the discriminating power of $\psi'$ becomes less pronounced overall, consistent with the smaller energy transfer of antineutrino interactions. As a result, the separation between model families is reduced compared with the neutrino case. Nevertheless, while the HF-CRPA distribution is close to SuSAv2 in the neutrino sample, it becomes more similar to the SF predictions in the antineutrino sample. This trend is consistent with the behavior already identified in the GENIE implementation of HF-CRPA, where the model approaches SuSAv2 at large momentum transfer but departs from it at lower energy and momentum transfer, where the genuine HF-CRPA response becomes more visible \cite{Dolan2022}. It is also worth noting that HF-CRPA and SuSAv2 are mean-field-inspired descriptions at the interaction-vertex level, whereas the construction of the hadronic final-state kinematics in the generator remains closer to an LFG-like picture \cite{Dolan2022, Dolan2020}. In this context, the overlap of the $\psi'$ distributions predicted by HF-CRPA, SuSAv2, and ED-RMF indicates a similar population of the superscaling phase space despite their different treatments of the hadronic final state. This agreement is compatible with an
important role of the vertex-level nuclear response.

\subsection{Observables from the \texorpdfstring{$\psi'$}{psi'} expansion: \texorpdfstring{$L_1$}{L1}, \texorpdfstring{$L_2$}{L2} and \texorpdfstring{$L_3$}{L3}}
\label{sec:L1_L2}
\subsubsection{\(L_1\) across nuclear models}
\label{subsubsec:L1_quasilinear}
\begin{figure}[b]
    \centering
        \includegraphics[width=0.85\linewidth]{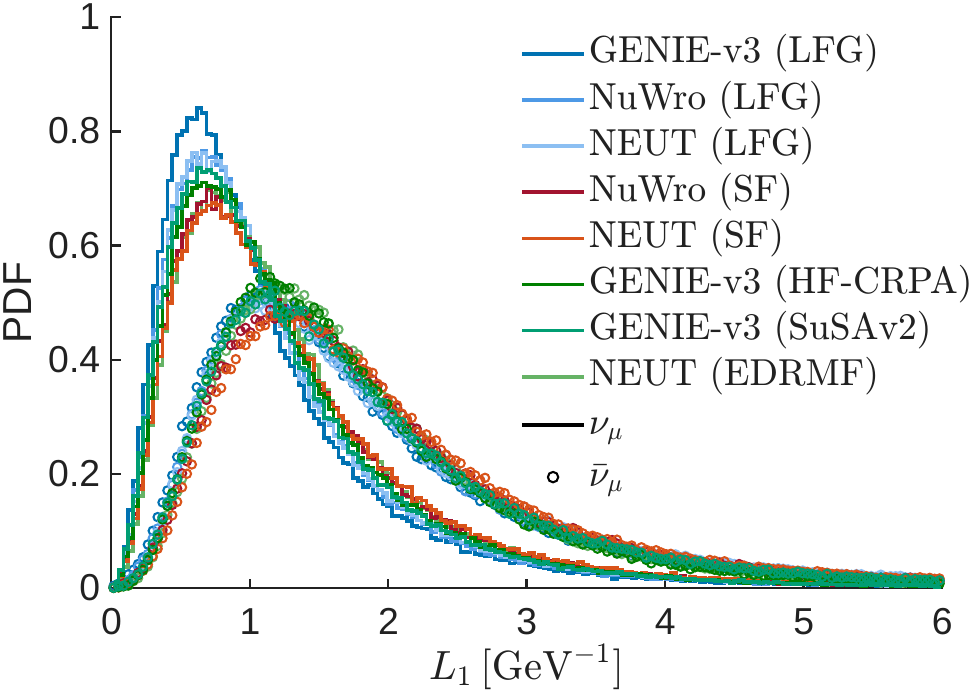}
   \caption{Muon-only coefficient $L_1(\vec p_\mu)$, shown as overlaid PDFs for all generator setups for $\nu_\mu$ and $\bar\nu_\mu$. Solid lines denote neutrinos, while circular markers denote antineutrinos.}
    \label{fig:L1overlay}
\end{figure}
As shown in Sec.~\ref{L1expl}, \(\psi'\) is well described by a linear dependence on the LQE energy-bias variable: \(\Delta E_\nu\) at truth level and \(\Delta E_p+S^{\rm eff}_{\rm RE}-\bar S_{\rm LQE}\) at reconstructed level.
To first order, the corresponding linear coefficient is \(L_1(\vec p_\mu)\), which encodes the local linear response of \(\psi'\) while depending only on the outgoing-muon kinematics and therefore being constructible from measured muon kinematics.

Figure~\ref{fig:L1overlay} shows overlaid probability density functions (PDFs) of $L_1$ for $\nu_\mu$ and $\bar\nu_\mu$.
The $L_1$ distributions differ across generator configurations, especially for $\nu_\mu$, reflecting differences in the predicted muon kinematics. These differences appear in both the peak positions and the tails of the \(L_1\) distributions. These model-dependent differences quantify how the different predictions populate the LQE quasi-linear region.
These differences manifest themselves as shifts in the \(L_1\) peak position, which sets the local linear response of $\psi'$ through the factorized form $\psi'\simeq \chi_F\,L_1 \times \bigl[(E_p-E_p^{\rm LQE})+\bigl(S^{\rm eff}_{\rm RE}-\bar S_{\rm LQE}\bigr)]$ in the LQE quasi-linear region.

Moreover, because \(L_1\) determines the upper bound on the neutrino-energy bias derived in Sec.~\ref{sec:L1DEnu}, changes in its median value and peak position directly affect the corresponding median upper bound. Table~\ref{tab:L1_L2_median_bounds} summarizes these bounds for all the models considered. Therefore, part of the model dependence originally encoded in the superscaling variable $\psi'$ is retained in the observable $L_1$ and, consequently, in the nuclear-model-dependent bound on $\Delta E_\nu$. This bound is discussed in detail in the following subsection.

A systematic neutrino--antineutrino difference is also visible in Fig.~\ref{fig:L1overlay}: the antineutrino samples populate larger values of $L_1$ than the corresponding neutrino samples, reflecting
the different muon kinematic regions populated by the two samples.
Through the first-order superscaling relation
$\Delta E_{\nu,\max}^{(1)}=1/(\chi_F L_1)$, this implies a more
restrictive constraint on the neutrino-energy reconstruction bias in
the antineutrino channel. This provides a superscaling interpretation of the narrower antineutrino reconstruction-bias distribution observed in Fig.~\ref{fig:REQE_compare}.

\subsubsection{$L_1$ as a nuclear bound on the neutrino-energy reconstruction bias}
\begin{figure}[b]
    \centering
    \includegraphics[width=0.6\linewidth]{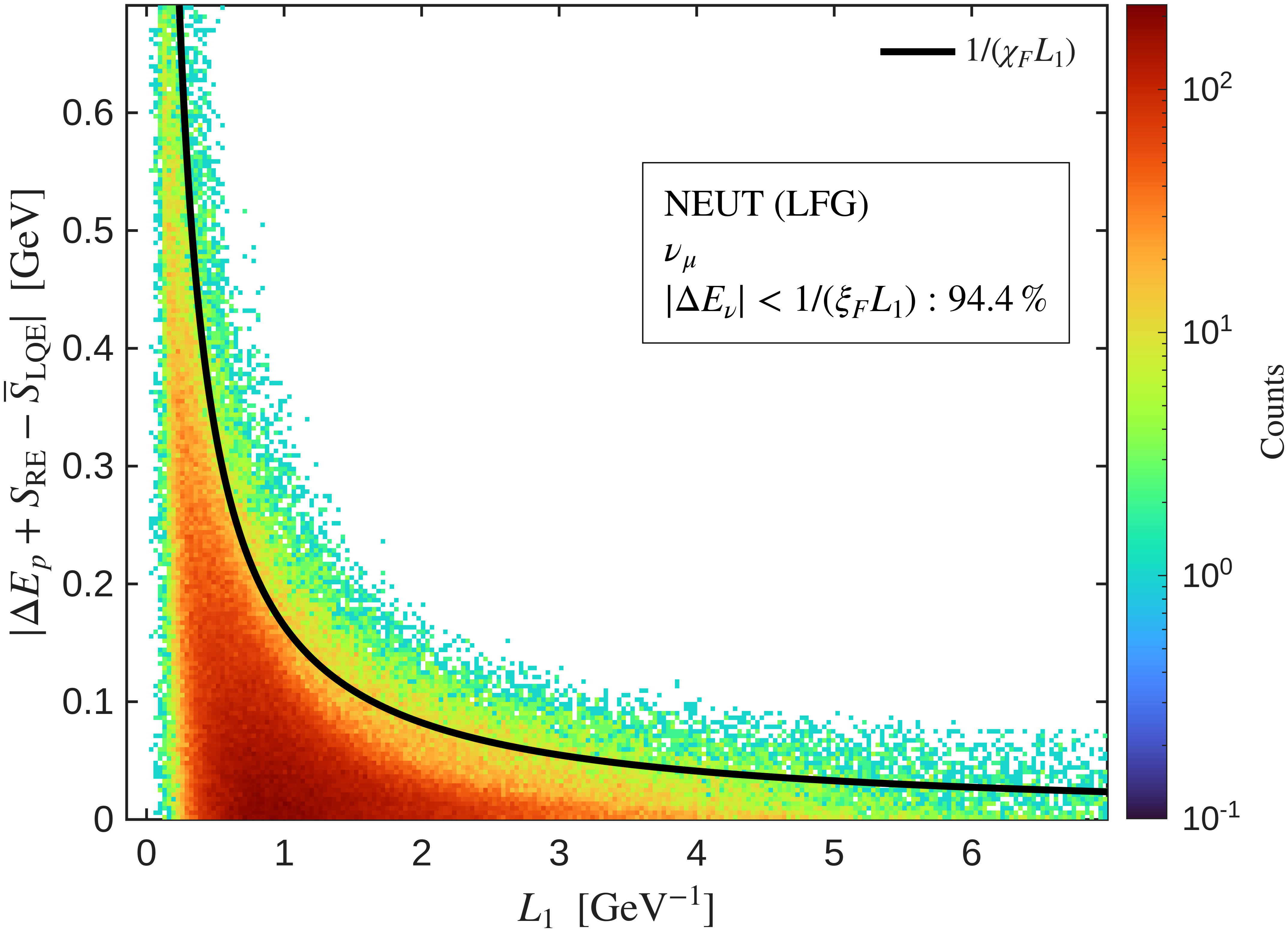}
    \includegraphics[width=0.6\linewidth]{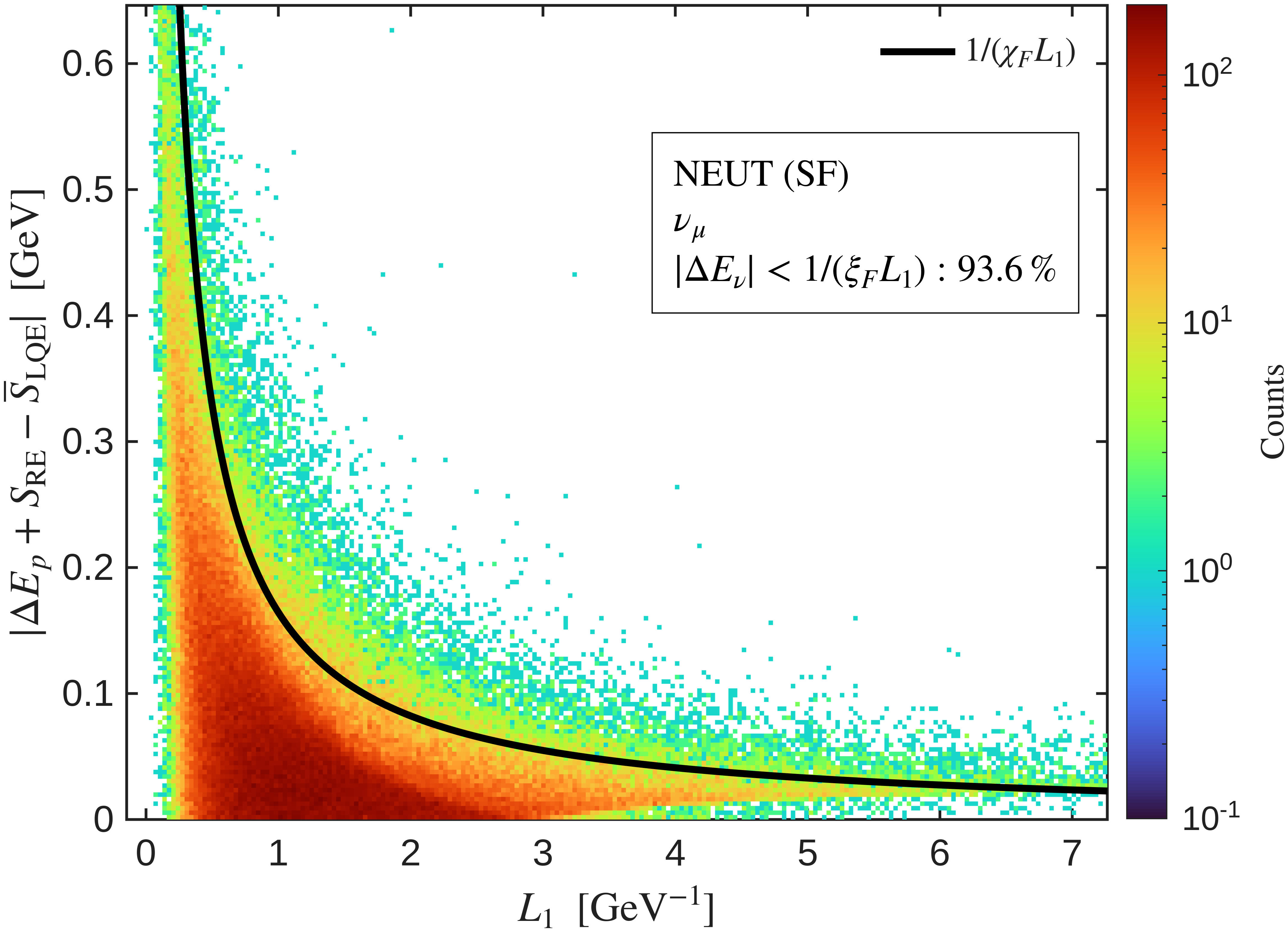}
    \includegraphics[width=0.6\linewidth]{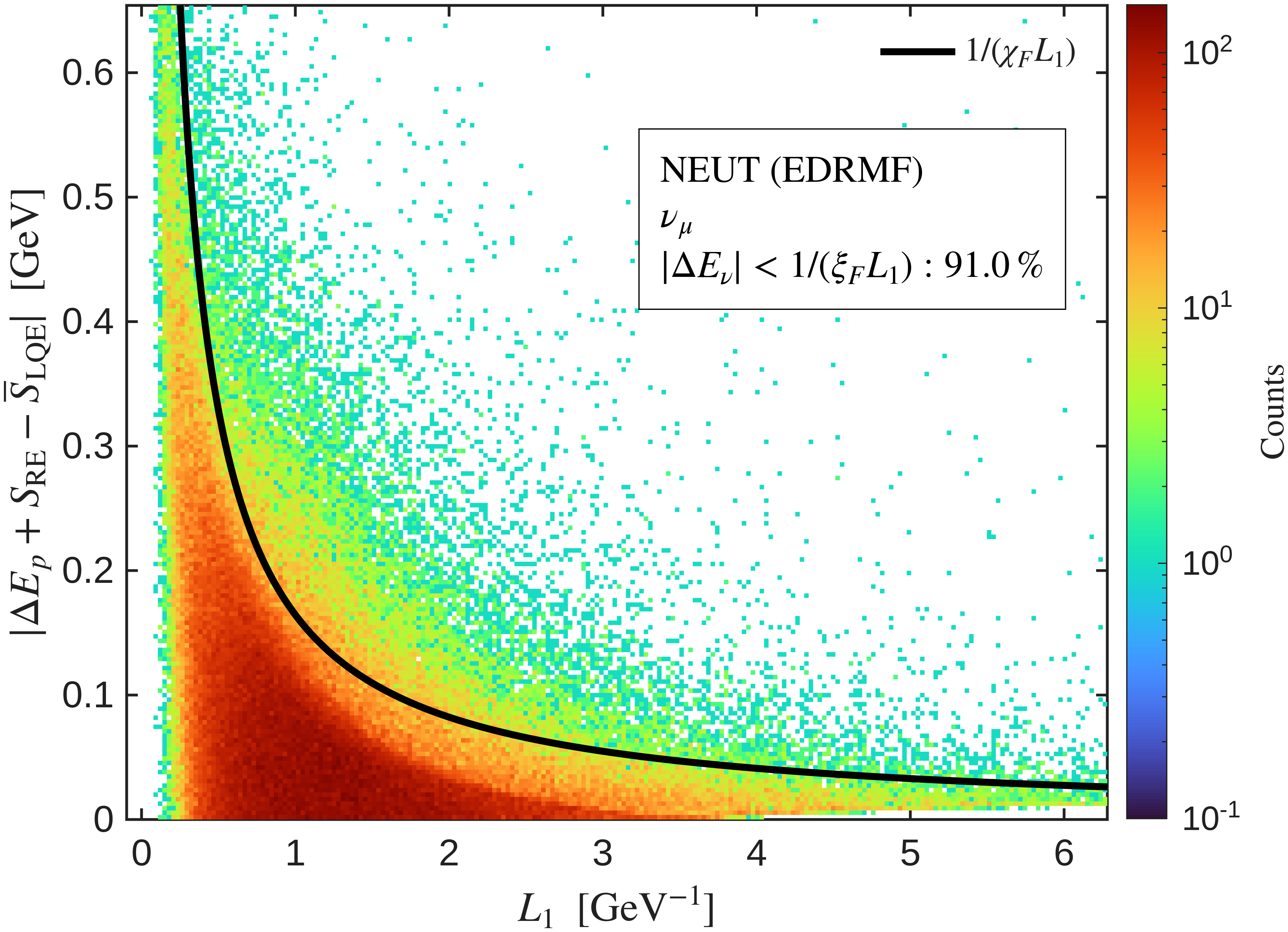}
    \caption{Correlation between the reconstructed neutrino-energy bias $|\Delta E_p+S^{\rm eff}_{\rm RE}- \bar S_{\rm LQE}|$ and the muon-only coefficient $L_1$ for the NEUT LFG, SF, and ED-RMF models. The black curve shows the first-order superscaling bound $1/(\chi_F L_1)$.}
    \label{fig:L1DEnumodels}
\end{figure}

We saw in the previous section that differences arise between nuclear models in the $L_1$ distributions. As shown in Sec.~\ref{sec:L1DEnu}, $L_1$ provides a nuclear bound on the neutrino-energy reconstruction bias. The differences between nuclear models reflect their different populations of the muon phase space and therefore of the event-by-event bound. In particular, SF models exhibit longer $\psi'$ tails outside the region $-1<\psi'<1$ than LFG models. Events in these tails lie outside the nominal region \(|\psi'|<1\) from which the bound is derived, and are therefore not expected to be contained by it. This effect is even more pronounced for the mean-field models shown in Fig.~\ref{fig:psiprime_reco_overlays}. The median values of the first-order bound, listed in the first and third columns of Table~\ref{tab:L1_L2_median_bounds}, display a clear model dependence. In the neutrino channel, the LFG models generally yield the largest median bounds, while the SF and ED-RMF models yield smaller values; HF-CRPA and SuSAv2 occupy an intermediate region. The corresponding differences are less pronounced in the antineutrino channel. Within our framework, the bound can be tested experimentally at a near detector by reconstructing the corresponding hadronic energy-bias variable,
\(\Delta E_p+S^{\rm eff}_{\rm RE}-\bar S_{\rm LQE}\), and could therefore be used to constrain nuclear models. Figure~\ref{fig:L1DEnumodels} illustrates this behavior through the correlation between the reconstructed neutrino-energy bias and $L_1$. The LFG sample remains strongly constrained by the analytical bound, whereas the SF and ED-RMF samples exhibit a larger population of events outside the nominal superscaling region.

\begin{table}[b]
\centering
\caption{Median first- and second-order neutrino-energy reconstruction bounds for $\nu_\mu$ and $\bar\nu_\mu$ samples, defined in Eqs.~\eqref{eq:DeltaEnu_bound_L1} and \eqref{eq:DeltaEnu_bound_L2}. All values are given in MeV.}
\label{tab:L1_L2_median_bounds}
\small
\setlength{\tabcolsep}{3pt}
\begin{tabular}{lcccc}
\hline
\\[-1.8ex]
& \multicolumn{2}{c}{$\Delta E_{\nu,\max}$}
& \multicolumn{2}{c}{$\Delta E_{\bar\nu,\max}$} \\
Model
& $(1)$
& $(2)$
& $(1)$
& $(2)$ \\
\hline
GENIE (LFG)      & $185$ & $158$ & $107$ & $96$ \\
NuWro (LFG)      & $172$ & $148$ & $104$ & $94$ \\
NEUT (LFG)       & $171$ & $147$ & $102$ & $91$ \\
NuWro (SF)       & $155$ & $134$ & $99$  & $89$ \\
NEUT (SF)        & $152$ & $132$ & $96$  & $86$ \\
GENIE (HF-CRPA)  & $162$ & $140$ & $107$ & $96$ \\
GENIE (SuSAv2)   & $167$ & $144$ & $104$ & $93$ \\
NEUT (ED-RMF)    & $154$ & $133$ & $104$ & $93$ \\
\hline
\end{tabular}
\end{table}

\subsubsection{\texorpdfstring{$L_2$}{L2} and \texorpdfstring{$L_3$}{L3}: curvature and sensitivity to nonlinear tails}
\label{subsubsec:L2_nonlinear}

In the LQE-linear region, $\psi'$ is dominated by the first-order term controlled by $L_1(\vec p_\mu)$ (Sec.~\ref{L1expl}). The second-order coefficient $L_2(\vec p_\mu)$ governs the leading departure from this linear behavior and therefore quantifies the local curvature of $\psi'$ as the energy-bias variable moves away from the quasielastic point. Close to the TAR ridge, the quadratic contribution $L_2(\Delta E_\nu)^2$ remains subleading, whereas it becomes increasingly relevant at larger reconstruction biases.

\begin{figure}[t]
    \centering
    \includegraphics[width=0.74\linewidth]{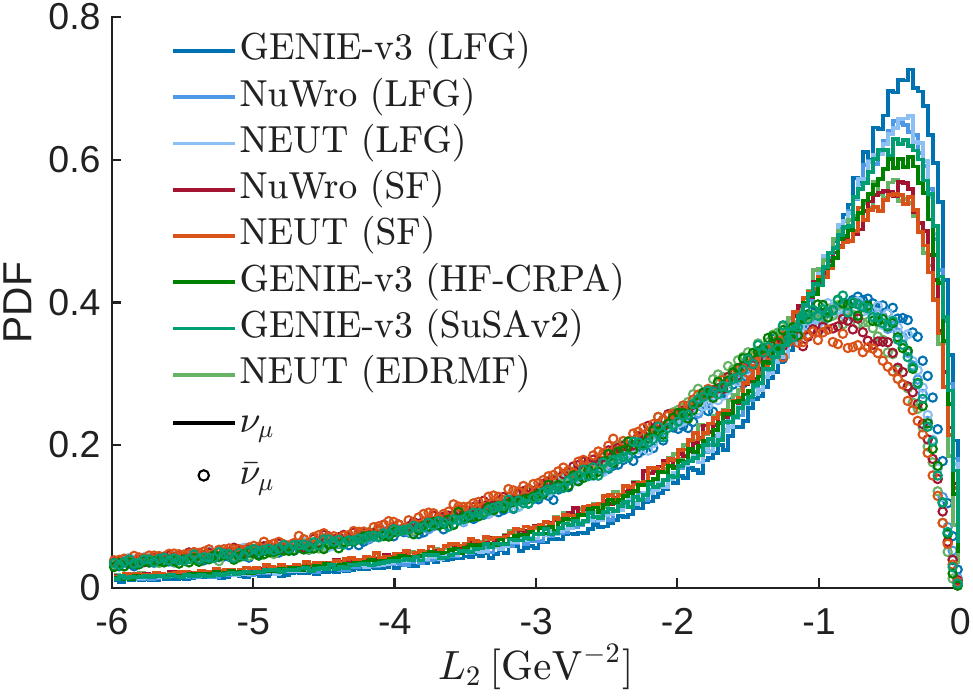}
    \includegraphics[width=0.74\linewidth]{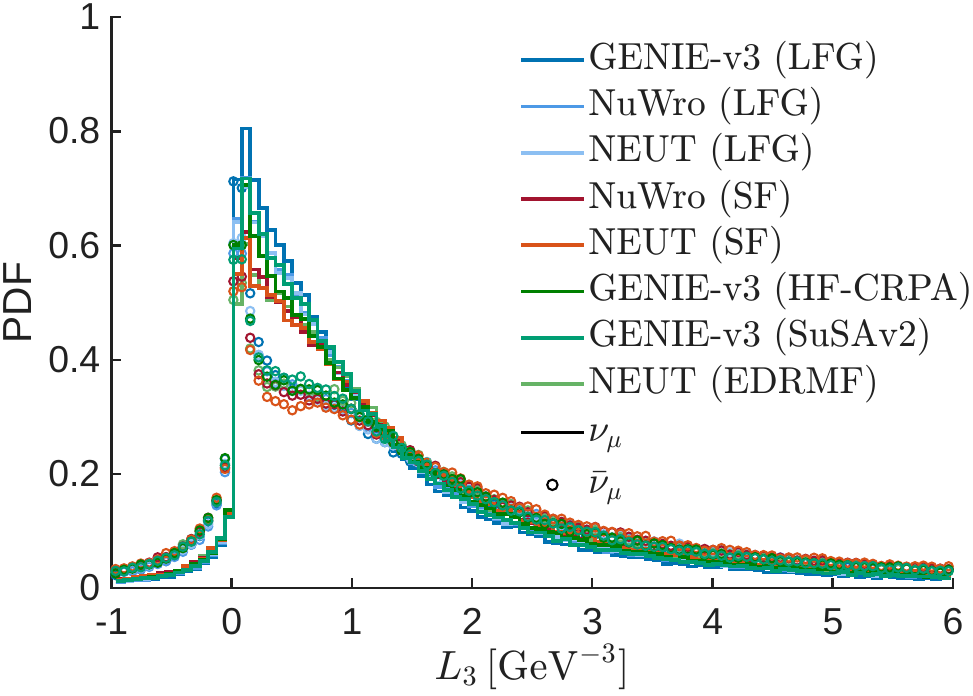}
    \caption{Muon-only coefficients \(L_2(\vec p_\mu)\) and \(L_3(\vec p_\mu)\), shown as overlaid PDFs for all generator setups. The upper plot shows \(L_2\) for \(\nu_\mu\) and \(\bar\nu_\mu\), while the lower plot shows the corresponding \(L_3\) distributions.}
    \label{fig:L2L3overlay}
\end{figure}

Figure~\ref{fig:L2L3overlay} shows the overlaid $L_2$ distributions for $\nu_\mu$ and $\bar\nu_\mu$. In the neutrino channel, the separation between nuclear models is clearly visible, and is at least comparable to that observed for $L_1$. Although $L_2$ depends only on the outgoing-muon kinematics, different nuclear descriptions populate the muon phase space differently, leading to distinct distributions across model families. In the antineutrino channel, this separation is significantly reduced, consistent with the weaker model dependence already observed for the other superscaling observables.

The model dependence carried by $L_2$ also enters directly into the second-order neutrino-energy reconstruction bound. The corresponding values are reported in the second and fourth columns of Table~\ref{tab:L1_L2_median_bounds} for neutrino and antineutrino samples, respectively. The spread between models is again more pronounced for neutrinos, while the antineutrino bounds remain more closely clustered.

\begin{figure}[b]
    \centering
    \includegraphics[width=0.7\linewidth]{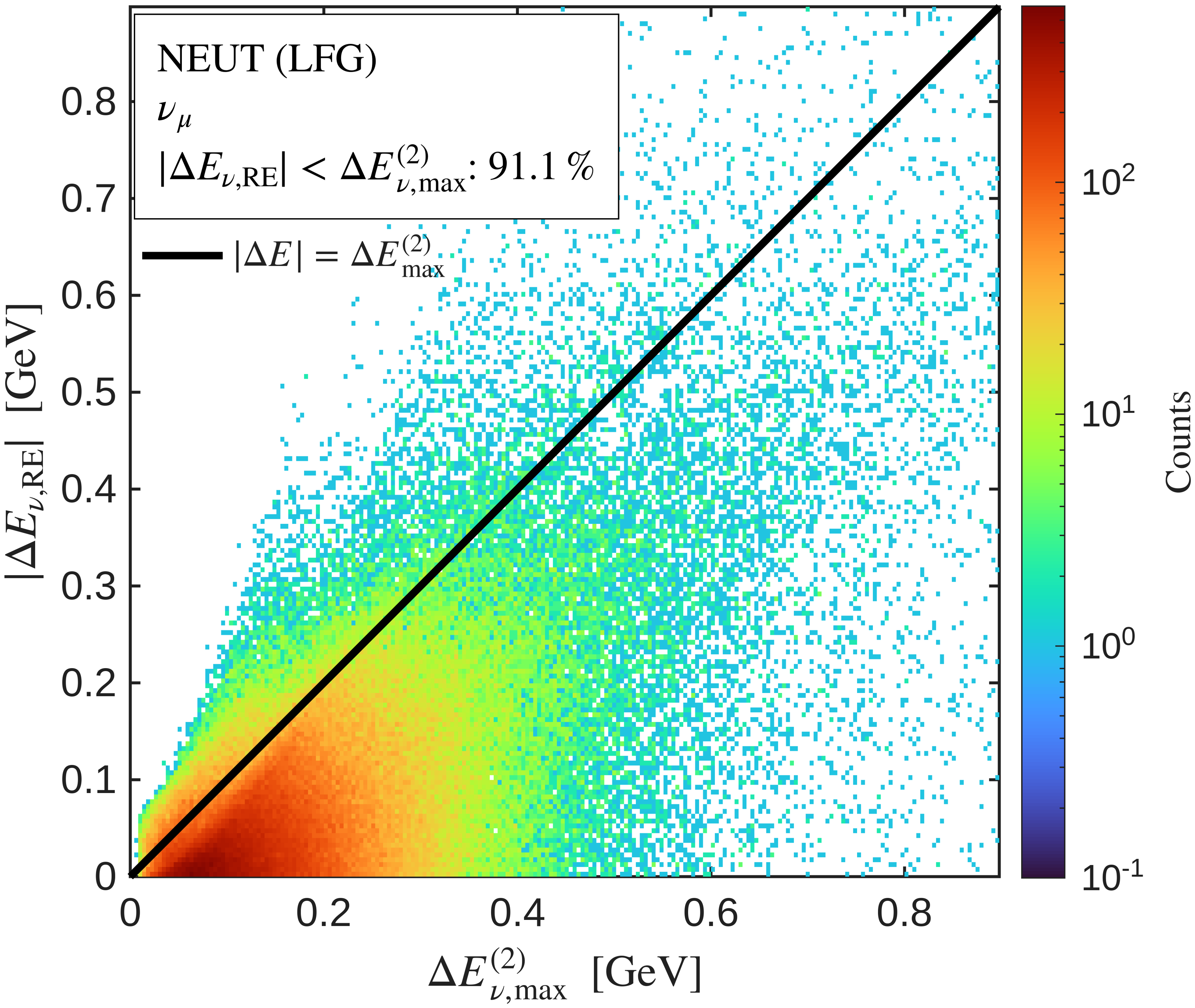}
    \caption{Event-by-event comparison between the reconstructed neutrino-energy bias \(\lvert\Delta E_{\nu,\mathrm{RE}}\rvert\) and the second-order superscaling bound \(\Delta E_{\nu,\max}^{(2)}\) for the NEUT LFG model. Events below the diagonal satisfy the quadratic constraint.}
    \label{fig:L2_bound_reco}
\end{figure}

As in the first-order case, the second-order constraint can be tested event by event at a near detector by comparing the reconstructed hadronic bias
\(\left|\Delta E_{\nu,\mathrm{RE}}\right|\)
with \(\Delta E_{\nu,\max}^{(2)}\).
Figure~\ref{fig:L2_bound_reco} shows this comparison for the NEUT LFG model. Most events lie below the diagonal, showing that the second-order analytical bound remains effective after reconstruction. Since \(L_1\), \(L_2\), and \(\Delta E_{\nu,\mathrm{RE}}\) are reconstructed from the outgoing-muon and proton kinematics, all quantities entering this test are experimentally accessible at a near detector.


The third-order coefficient $L_3(\vec p_\mu)$ controls the next correction in the Taylor expansion: it encodes the cubic deformation of $\psi'$ as the energy-bias variable moves farther away from the LQE point.
Accordingly, while the cubic contribution remains subleading close to the LQE point, large values of \(|L_3|\) identify kinematic regions in which the third-order correction can become relevant once \(|\Delta E_\nu|\) is sufficiently large.

Figure~\ref{fig:L2L3overlay} shows the overlaid PDFs of $L_3$ for $\nu_\mu$ and $\bar\nu_\mu$.
In contrast to $L_1$ and $L_2$, the $L_3$ distributions develop a high-value tail, driven by the same forward enhancement in the $p_{T\mu}\to0$ limit.
As a result, $L_3$ is significantly less stable as an event-by-event observable: its distribution is more sensitive to rare kinematic configurations with very small muon transverse momentum, making a detailed comparison across nuclear models less robust and less informative than for the lower-order coefficients.

\label{subsubsec:L1L2_model_spread}
\section{Stability of the superscaling observables across flux and target configurations}
\label{sec:exp_outlook}
The configurations considered combine the T2K near-detector (ND), NuMI low-energy (LE), and NuMI medium-energy (ME) fluxes with a carbon-based target, and the DUNE ND, MicroBooNE, and SciBooNE fluxes with an argon-based target~\cite{ABE2011106,ADAMSON2016279,DUNE:2015lol,Acciarri2017,SciBooNE:2006asj}. They should be understood as generated flux-target configurations rather than exact reproductions of all the corresponding experimental detector materials.

For the computation of the superscaling observables, we use a single target-dependent reference value for all configurations: \(\bar S_{\rm LQE}=29~\mathrm{MeV}\) for carbon and \(\bar S_{\rm LQE}=18~\mathrm{MeV}\) for argon. These values correspond to the averages of the peak positions of $S_{\rm LQE}$, whose distribution is defined in Eq.~(\ref{SQE}), obtained for the configurations associated with each target. The individual peak values are reported in Appendix~\ref{Sec:ExperimentalConditions}. The variations of the \(S_{\rm LQE}\) peak positions within each
target group remain relatively small, motivating the use of a common
target-dependent reference value. The larger variations affecting the
hadronic reconstruction are instead accounted for through the tuned
parameter \(S_{\rm RE}^{\rm eff}\). For the present comparison, we keep $k_F = 228$~MeV for the carbon-based configurations and adopt $k_F = 241$~MeV for the argon-based ones \cite{Barbaro2019}. 

\begin{figure}[t]
    \centering
    \includegraphics[width=0.75\linewidth]{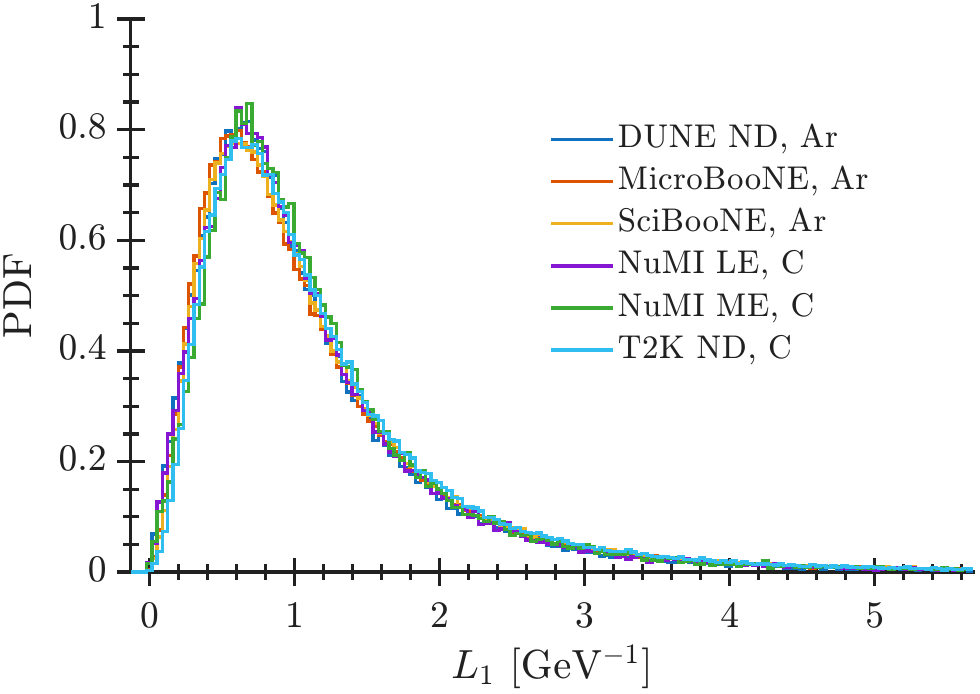}
     \includegraphics[width=0.75\linewidth]{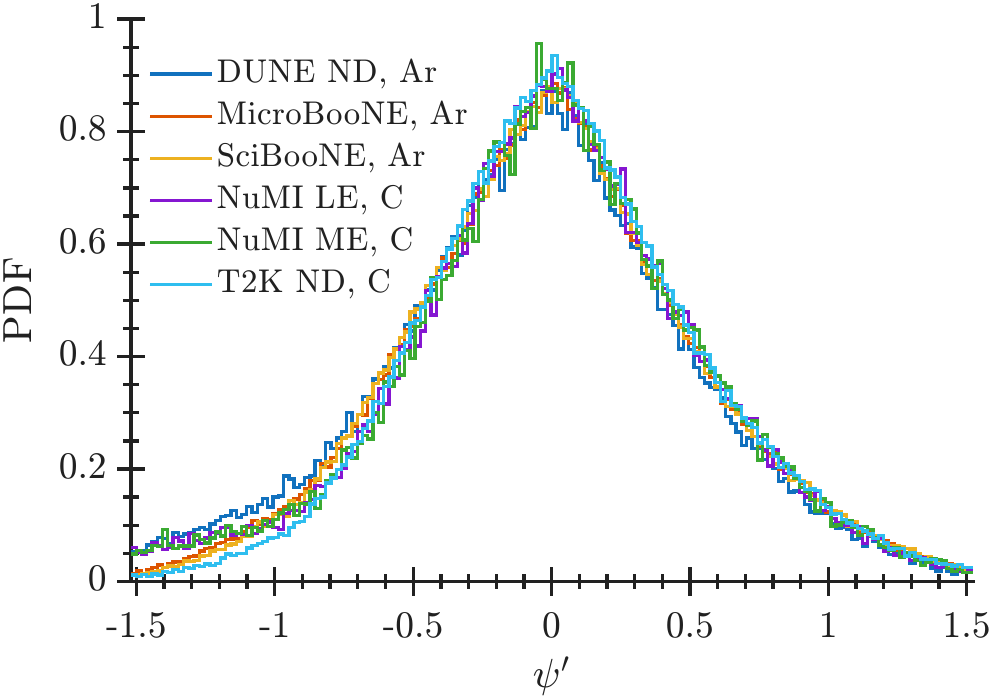}
    \caption{Comparison of $L_1$ (top) and tuned $\psi'$ (bottom) distributions for several generated flux--target configurations with different neutrino fluxes and target materials: T2K ND, NuMI ME, and NuMI LE on carbon, and MicroBooNE, SciBooNE, and DUNE ND on argon. To isolate the dependence on the flux and target, all samples are generated with the same theoretical model, the NEUT LFG model.}
    \label{figL1psiprime_exp}
\end{figure}

As expected, the stability of $\psi'$ is supported by the overlap of the corresponding distributions across the different flux--target configurations (see Fig.~\ref{figL1psiprime_exp}), within the superscaling region where scaling is expected to hold, namely for $-1 \lesssim \psi' \lesssim 1$. This shows that, once properly calibrated, $\psi'$ remains largely insensitive to the details of the incoming flux and target environment.

A comparable stability is also observed for the muon-only observable $L_1$, as evidenced by the overlap of the $L_1$ distributions across the different flux--target configurations shown in Fig.~\ref{figL1psiprime_exp}. The stability of \(L_1\) has a direct interpretation within the
present framework. Through \(\Delta E_{\nu,\max}^{(1)}=1/(\chi_F L_1)\), \(L_1\) characterizes, event by event, the nuclear envelope of the kinematic region associated with \(|\psi'|<1\). This envelope is set
by the superscaling condition and the corresponding target-dependent
nuclear parameters. Notably, the different kinematic distributions shown in
Fig.~\ref{fig:experiemnt_kin} map onto very similar normalized
\(L_1\) distributions in Fig.~\ref{figL1psiprime_exp}. This indicates
that, once the target-dependent superscaling parameters are accounted
for, the constraint encoded by \(L_1\) is governed by the nuclear superscaling envelope and is only weakly affected by the
incident-flux energy scale. 

This property is important for experimental applications.
\(L_1\) is a purely muonic observable shown to remain approximately stable under simultaneous changes of the incident neutrino flux and nuclear target. This opens the possibility of measuring \(L_1\) across different detector configurations and directly comparing the resulting distributions. After accounting for the acceptance, resolution, and systematic uncertainties specific to each detector, compatible measurements are therefore expected to be obtained across experiments probing different neutrino
fluxes and nuclear targets.

\section{Conclusion}
\label{sec:conclusions}

In this work, we developed a unified framework for neutrino-energy reconstruction that connects two complementary approaches: a removal-energy reconstruction using both the outgoing-muon and hadronic kinematics, and a lepton-only quasielastic reconstruction based exclusively on the outgoing-muon kinematics. This unification led to an identity relating the reconstructed neutrino-energy bias,
\(\Delta E^{\rm RE}_\nu=E^{\rm RE}_\nu-E_\nu^{\rm LQE}\),
to the reconstructed hadronic combination
\(\Delta E_p+S_{\rm RE}-\bar S_{\rm LQE}\). We then used this framework to reparametrize the superscaling variable \(\psi'\) in terms of the neutrino-energy reconstruction bias, thereby establishing a direct analytical connection between superscaling and neutrino-energy reconstruction. A Taylor expansion around the quasielastic target-at-rest point was then performed to factorize the dependence on the neutrino-energy bias from a hierarchy of coefficients \(L_n(\vec p_\mu)\) that depend only on the outgoing-muon kinematics. This factorization separates the energy-bias variable, which is not directly accessible without hadronic information, from a set of fully muonic observables that can be reconstructed event by event. The superscaling condition \(|\psi'|<1\) allowed us to derive the event-by-event first-order bound
\[
|\Delta E_\nu|<
\Delta E_{\nu,\max}^{(1)}
=
\frac{1}{\chi_FL_1},
\]
which depends only on the outgoing-muon kinematics and the target-dependent superscaling parameters. Including the quadratic coefficient \(L_2\) provides a tighter second-order refinement of this constraint. We then performed a systematic comparison of nuclear-model predictions using CC1p1h samples generated with GENIE, NEUT, and NuWro, including local-Fermi-gas, spectral-function, and mean-field-based descriptions. After calibrating the global \(\psi'\) peak position through \(S^{\rm eff}_{\rm RE}\), the residual distributions were found to cluster primarily according to the nuclear-model family rather than the generator implementation, demonstrating the nuclear model sensitivity of both \(\psi'\) and the coefficients \(L_n\). The resulting correlations reveal model-dependent differences in the coverage of the analytical envelopes, with LFG samples remaining more strongly confined to the nominal superscaling region than SF and mean-field-based samples. Finally, keeping the nuclear-interaction model fixed, we varied the incident neutrino flux and target material to simulate several flux--target configurations. After adjusting the target-dependent superscaling parameters, the reconstructed \(\psi'\) distributions exhibit the expected approximate stability across these configurations. A comparable stability is also observed for \(L_1\), despite the substantially different kinematic distributions produced by the various fluxes. Within the present framework, this behavior can be interpreted through the role of \(L_1\) in characterizing the nuclear superscaling
envelope of the allowed reconstruction bias. Unlike the reconstructed \(\psi'\), \(L_1\) requires no outgoing-nucleon information and is therefore accessible in antineutrino samples and at far detectors where the neutrino energy is reconstructed from the outgoing-lepton kinematics alone.
The observable \(L_1\) could therefore provide a new event-by-event diagnostic of neutrino-energy reconstruction quality in oscillation experiments, grounded directly in the superscaling properties of the nuclear response. The systematically larger \(L_1\) values populated by the antineutrino samples lead to tighter superscaling bounds and provide an interpretation of their narrower neutrino-energy reconstruction
bias distributions. Because \(L_1\) characterizes the target-dependent superscaling constraint and its distribution is observed to remain approximately stable across the flux--target configurations considered, it could serve as a common reference observable for comparisons between different experimental configurations. Measurements of the \(L_1\) distribution and of its correlation with the reconstructed hadronic energy bias at near detectors could therefore provide new constraints on nuclear-model uncertainties relevant to oscillation analyses.

\begin{acknowledgments}
This work was supported by the Swiss National Science Foundation (SNSF) through the MAPS project (Project No.~230193, 01.07.2025--30.06.2029) at the University of Geneva. S.~Samani acknowledges support from the SNSF through a Swiss Postdoctoral Fellowship, under Grant No.~\path{TMPFP2_234635}. The authors are grateful to S.~Dolan, C.~Wilkinson, and J.~McKean for generating and providing the NUISANCE Monte Carlo files used in this analysis.
\end{acknowledgments}

\appendix

\section{$\psi'$ and the Fermi Sphere}
\label{app:psi_RFG}

Within the relativistic Fermi gas (RFG) model, the nuclear ground state is described by a uniform filling of momentum space up to the Fermi momentum \(k_F\). Nucleon momenta \(\vec p\) therefore satisfy
\begin{align}
|\vec p|
&\leq k_F,
\qquad
p_\parallel \equiv \vec p\cdot\hat q,
\\
p_T^2
&=
|\vec p|^2-p_\parallel^2
\leq
k_F^2-p_\parallel^2,
\end{align}
where
\(\hat q\equiv\vec q/|\vec q|\).
For fixed energy and momentum transfers \((\omega,|\vec q|)\), quasielastic kinematics primarily constrains the longitudinal component of the initial nucleon momentum, while the transverse components enter through the available phase space. Geometrically, for a fixed value of \(p_\parallel\), the allowed transverse momenta lie within a disk of radius
\(\sqrt{k_F^2-p_\parallel^2}\),
whose area is
\begin{align}
A(p_\parallel)
=
\pi\left(k_F^2-p_\parallel^2\right).
\end{align}

This geometrical construction provides an intuitive interpretation of the RFG scaling function. It is useful to introduce the reduced longitudinal coordinate
\begin{align}
u
\equiv
\frac{p_\parallel}{k_F},
\qquad
-1\leq u\leq1.
\end{align}
At fixed \(u\), the available transverse phase space is proportional to
\begin{align}
A(u)
=
\pi k_F^2\left(1-u^2\right).
\end{align}
In the RFG, the superscaling variable \(\psi'\) provides the corresponding relativistic scaling coordinate, expressing the minimum longitudinal momentum required by the scattering kinematics in units of the characteristic Fermi scale. In the geometrical limit, it is therefore naturally associated with the reduced coordinate \(u\).

After normalization, the transverse phase-space factor yields the well-known RFG scaling function
\begin{align}
f_{\mathrm{RFG}}(\psi')
=
\frac{3}{4}
\left(1-\psi'^2\right)
\Theta\!\left(1-|\psi'|\right),
\label{eq:fRFG}
\end{align}
with
\begin{align}
\int_{-1}^{1}
f_{\mathrm{RFG}}(\psi')\,d\psi'
=
1.
\end{align}
The point \(\psi'=0\) corresponds to the quasielastic peak and to a vanishing minimum longitudinal momentum, while \(|\psi'|=1\) marks the boundary of the ideal RFG phase space. This geometrical boundary motivates the use of the condition \(|\psi'|<1\) in deriving the neutrino-energy reconstruction constraints of Sec.~\ref{sec:L1DEnu}.

Nuclear models beyond the RFG predict scaling functions that depart from this symmetric inverted-parabolic form. These deviations reflect the underlying nuclear dynamics and the structure of the initial-state momentum and energy distributions. In particular, models with extended high-momentum components can populate the region \(|\psi'|>1\), which is strictly excluded in the ideal RFG.

The RFG scaling function \(f_{\mathrm{RFG}}(\psi')\) does not depend explicitly on \(\omega\) or \(|\vec q|\); the kinematic dependence enters only through the single variable \(\psi'(\omega,|\vec q|)\), embodying \emph{scaling of the first kind}. In the impulse approximation, the inclusive response can be factorized schematically into a single-nucleon contribution and a nuclear scaling function. For a suitably reduced response, or reduced cross section, one may write
\begin{align}
\frac{d^2\sigma_A}
{d\Omega_\ell\,dE_\ell}
\simeq
\sigma_{\ell N}(\omega,|\vec q|)
\,f(\psi'),
\label{eq:factor_scaling}
\end{align}
where \(\sigma_{\ell N}\) contains the elementary single-nucleon response and the relevant leptonic and kinematic factors, while \(f(\psi')\) encodes the nuclear many-body dynamics.

\section{Nuclear Models}
\subsection{Removal energy across nuclear models}
\label{App:Nuclearmodel}
We have explored various models available within the NUISANCE software package~\cite{Stowell:2016jfr}, as described in Section~\ref{Sec:NuclearModels}. 
However, these models differ significantly in their treatment of the nuclear ground state. Specifically, their descriptions of the interacting nucleon's kinematics and the removal energy vary considerably. The figures in this subsection focus on the CC1p1h neutrino channel on carbon, while Tables~\ref{tab:sre_mean} and~\ref{tab:sqe_peak} also report the corresponding antineutrino values used in the main analysis.

\begin{figure}[b]
    \centering    \includegraphics[width=0.45\textwidth]{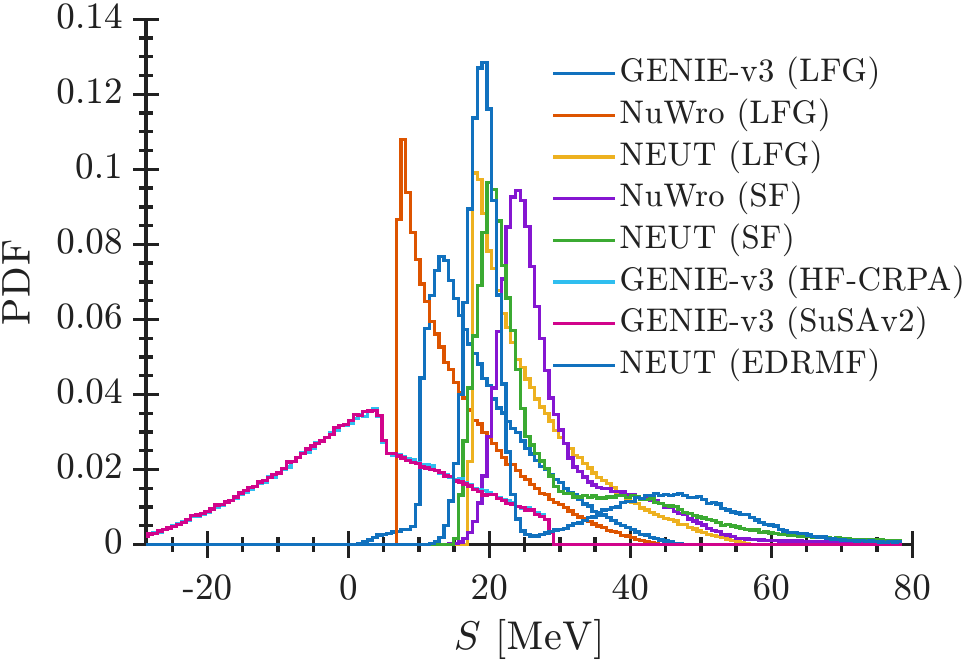}
  \caption{Removal-energy spectra for different nuclear-interaction models generated with the T2K ND FHC $\nu_\mu$ flux on a carbon target. The distributions are shown at truth level without NrS.}
    \label{fig:SRE_RemEn}
\end{figure}
\begin{figure*}[t]
    \centering
    \includegraphics[width=0.49\textwidth]{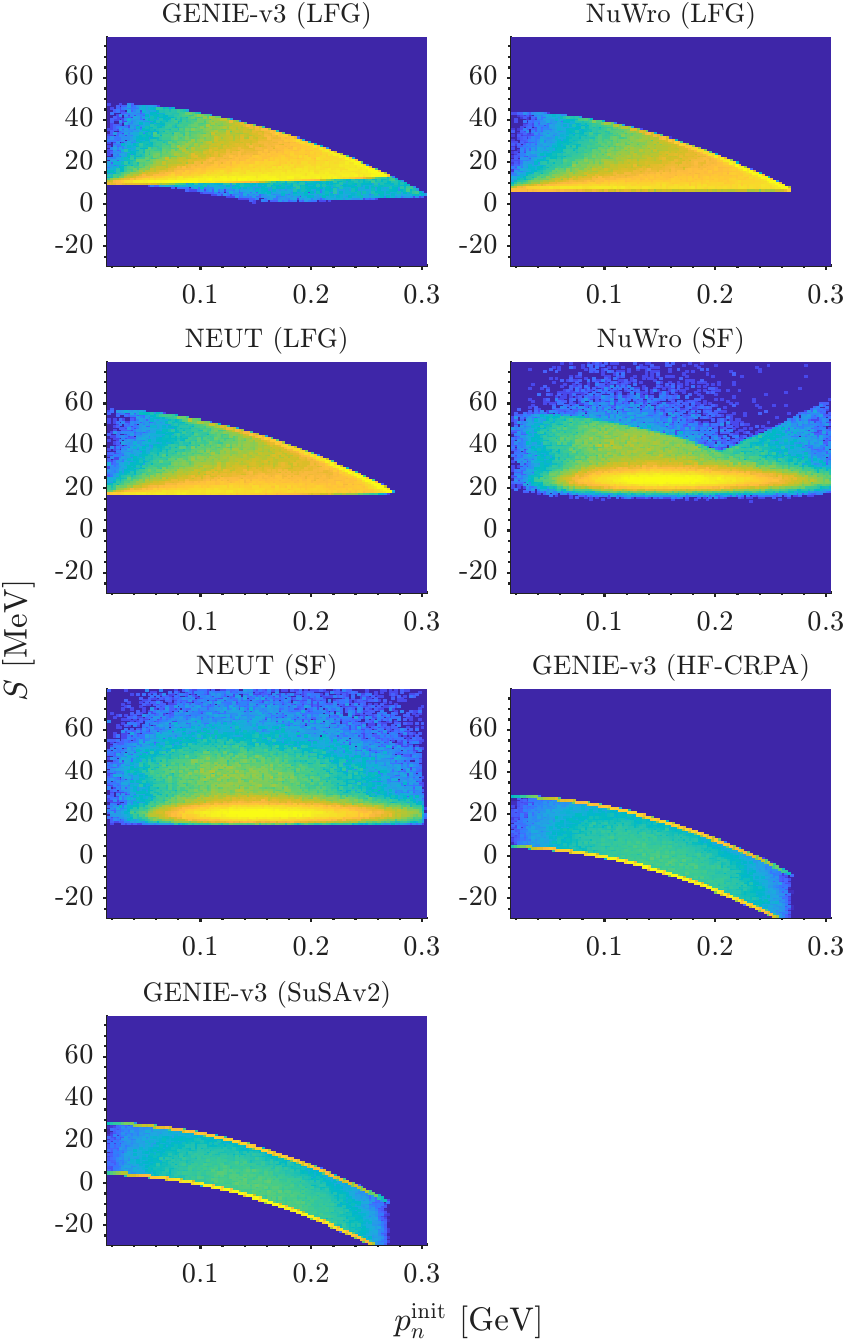}
        \includegraphics[width=0.49\textwidth]{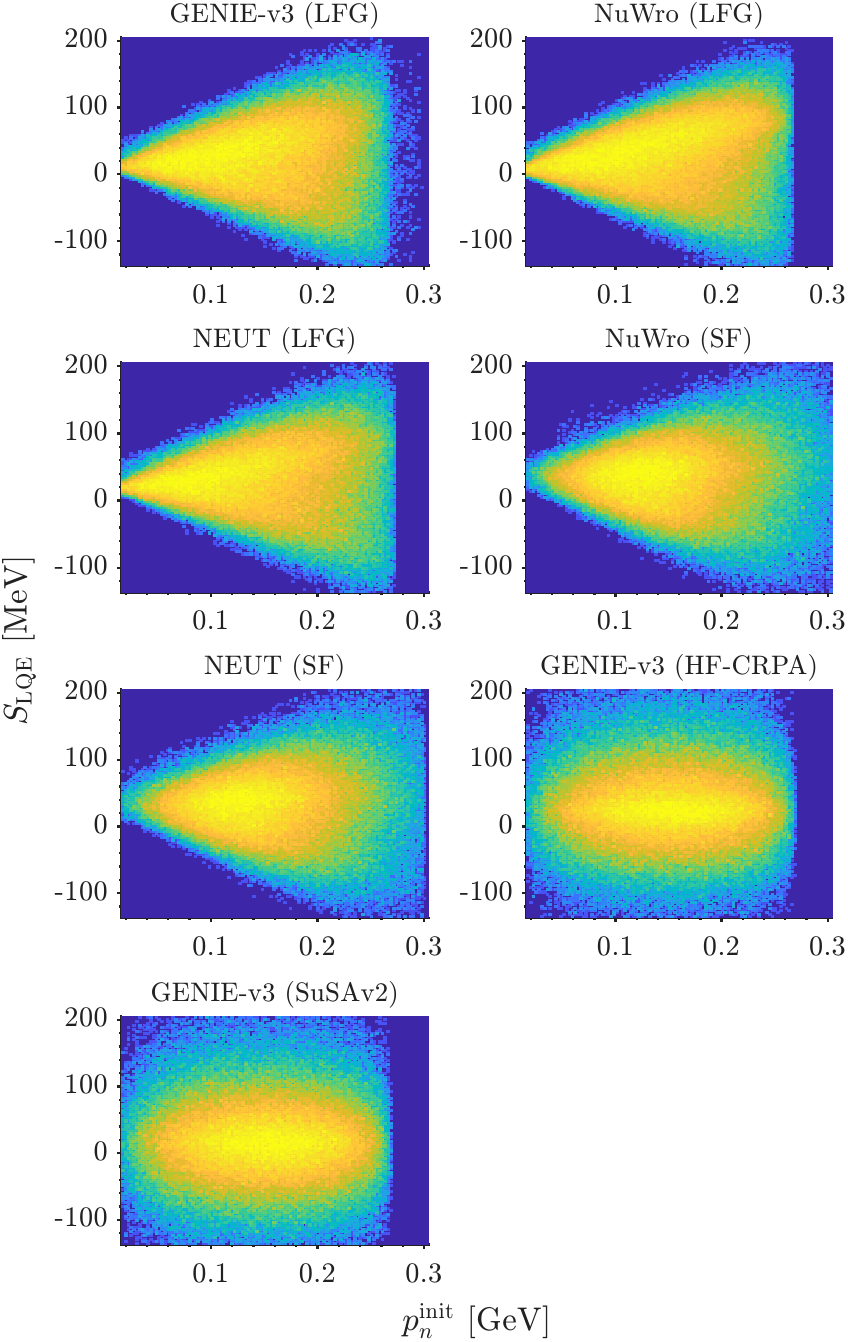}
    \caption{
\textbf{Left column:} Two-dimensional density distributions of the true removal energy, \(S = \omega - E_p + M_n\), versus the initial momentum of the target neutron, \(p^{\rm init}_n\), for the different generator configurations considered in this work. 
\textbf{Right column:} Two-dimensional density distributions of the removal energy built upon
the target-at-rest assumption,
$S_{\rm LQE} = \omega - \sqrt{M_p^2 + |\vec q|^2} + M_n$,
versus the initial nucleon momentum $p^{\rm init}_n$ for the same generator
configurations.
In both sets of plots, events are selected from the CCQE \(1p1h\) channel on a carbon target, with only the neutrino mode considered. }
    \label{fig:SRE_pinitN_matrix}
    \label{fig:dTAR_pinitN_matrix}
\end{figure*}

Figure \ref{fig:SRE_RemEn} presents the removal-energy spectra simulated for a carbon target under the T2K Near Detector forward horn current (FHC) $\nu_{\mu}$ flux. The distributions, shown at truth level without nuclear rescattering (NrS), illustrate the variations arising from different nuclear-interaction models.

The left column of Figure~\ref{fig:SRE_pinitN_matrix} displays the two-dimensional density distributions of the true removal energy ($S = \omega - E_p + M_n$) versus the initial momentum of the target neutron ($p^{\rm init}_n$) for the various generator configurations studied. The NEUT ED-RMF configuration is omitted from this comparison
because $p^{\rm init}_n$ is not available as an event-level
generator variable in the sample used here.

Finally, the right column of Figure~\ref{fig:dTAR_pinitN_matrix} presents the
two-dimensional density distributions of the removal-energy estimator built
upon the target-at-rest assumption,
$S_{\rm LQE} = \omega - \sqrt{M_p^2 + |\vec q|^2} + M_n$, as a function of
the initial neutron momentum $p^{\rm init}_n$ for the different
generator configurations considered in this work. This observable provides a
sensitive probe of the nuclear-model assumptions, particularly highlighting
how different treatments of the initial nucleon momentum and energy balance
affect the reconstructed quasielastic kinematics. 
The standard estimator $S_{\rm RE}$ exhibits a significant model dependence. 

The generator-level constructions used for GENIE HF-CRPA and SuSAv2
employ a different treatment of the initial-nucleon energy and do not
enforce the same event-by-event energy-balance relation as the LFG and
SF implementations. Consequently, the correlation between \(S\) and \(p_n^{\rm init}\) is not described by a simple Fermi-gas-like structure, and the width of \(S_{\rm LQE}\) cannot be interpreted solely as the progressive breakdown of the target-at-rest approximation with increasing initial momentum. As discussed in Ref.~\cite{Dolan2022}, the GENIE HF-CRPA implementation does not exactly enforce energy conservation event by event. The closely related construction used for SuSAv2 accounts for the similarly non-conical structure observed in that configuration.
\begin{table}[t]
\centering
\caption{Standard removal-energy estimator $S_{\rm RE}=\bar S $ for the $\nu$ and $\bar{\nu}$ samples of each generator configuration. The values are computed from vertex-level kinematics before nuclear rescattering and are given in MeV.}
\begin{tabular}{lcc}
\hline
\\[-1.8ex]
Model & $S_{\rm RE}^{\nu}$ [MeV] & $S_{\rm RE}^{\bar\nu}$ [MeV] \\
\hline
GENIE (LFG)      & $19$ & $19$ \\
NuWro (LFG)      & $16$ & $15$ \\
NEUT (LFG)       & $27$ & $26$ \\
NuWro (SF)       & $31$ & $22$ \\
NEUT (SF)        & $30$ & $27$ \\
GENIE (HF-CRPA)  & $2$  & $-3$ \\
GENIE (SuSAv2)   & $2$  & $-4$ \\
NEUT (ED-RMF)    & $29$ & $22$ \\
\hline
Average          & $19$ & $16$ \\
\hline
\end{tabular}
\label{tab:sre_mean}
\end{table}

\begin{table}[t]
\centering
\caption{Peak positions of the truth-level \(S_{\rm LQE}\) distributions for the \(\nu_\mu\) and \(\bar\nu_\mu\) samples of each generator configuration. Unless otherwise stated, we adopt the common reference value \(\bar S_{\rm LQE}=26\,\mathrm{MeV}\).}
\label{tab:sqe_peak}
\begin{tabular}{lcc}
\hline
\rule{0pt}{2.8ex}Model & $S_{\rm LQE,\mathrm{peak}}^{\nu}$ [MeV] & $S_{\rm LQE,\mathrm{peak}}^{\bar\nu}$ [MeV] \\
\hline
GENIE (LFG)      & $20$ & $29$ \\
NuWro (LFG)      & $22$ & $28$ \\
NEUT (LFG)       & $26$ & $32$ \\
NuWro (SF)       & $42$ & $36$ \\
NEUT (SF)        & $40$ & $39$ \\
GENIE (HF-CRPA)  & $23$ & $17$ \\
GENIE (SuSAv2)   & $16$ & $16$ \\
NEUT (ED-RMF)    & $19$ & $13$ \\
\hline
Model average    & $26$ & $26$ \\
\hline
\end{tabular}
\end{table}

The peak positions of the \(S_{\rm LQE}\) distributions exhibit a sizable model dependence, ranging from approximately \(13\) to \(42~\mathrm{MeV}\) across the neutrino and antineutrino samples. Despite this spread, the model-averaged peak positions are close to \(26~\mathrm{MeV}\) in both channels, motivating the common reference value \(\bar S_{\rm LQE}=26~\mathrm{MeV}\) adopted in the main analysis. The NuWro and NEUT SF configurations favor the largest values, while GENIE SuSAv2 gives one of the lowest estimates.

\section{Kinematic distributions across flux--target configurations}
\label{Sec:ExperimentalConditions}
\begin{figure}[H]
    \centering
    \includegraphics[width=0.450\linewidth]{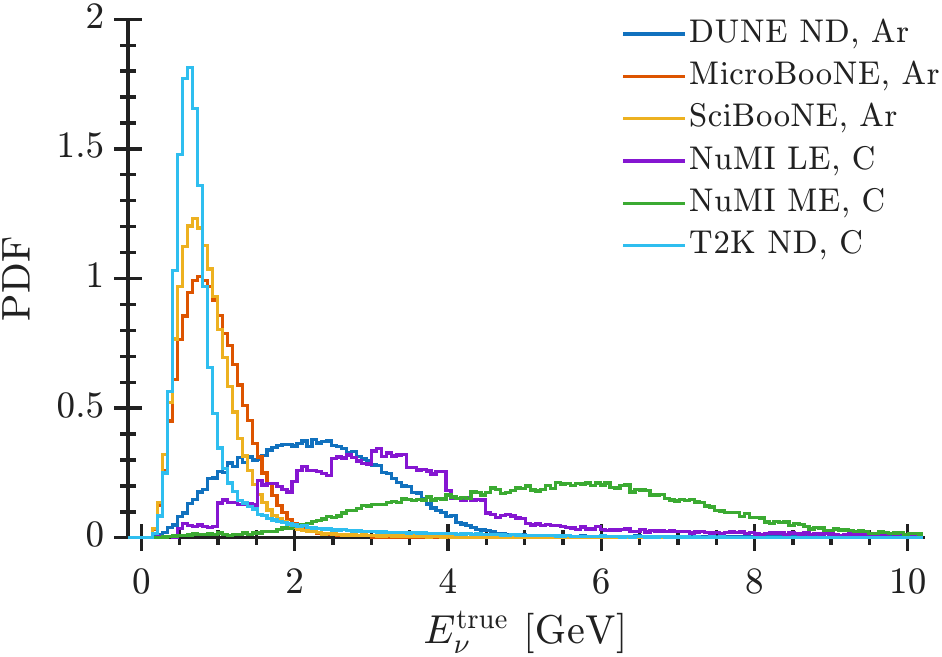}
    \hfill
    \includegraphics[width=0.450\linewidth]{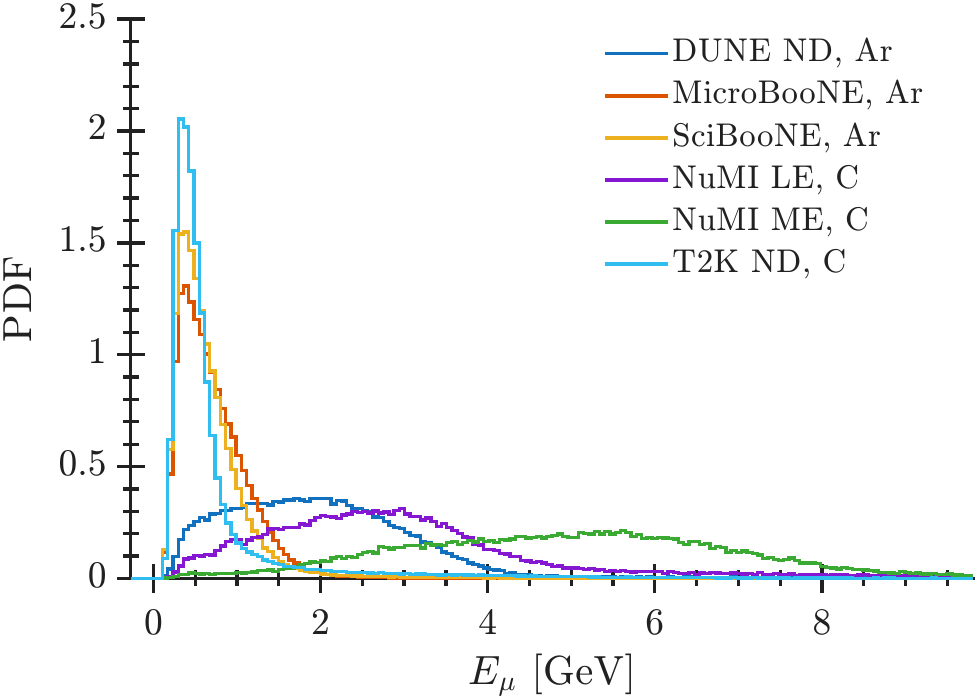}

    \includegraphics[width=0.45\linewidth]{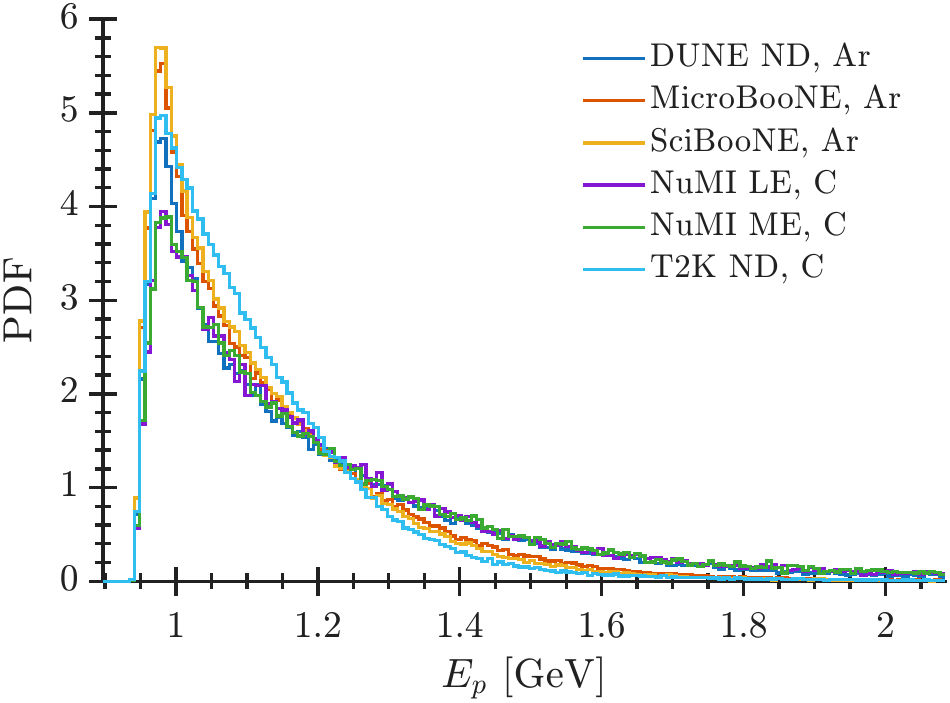}
    \hfill
    \includegraphics[width=0.45\linewidth]{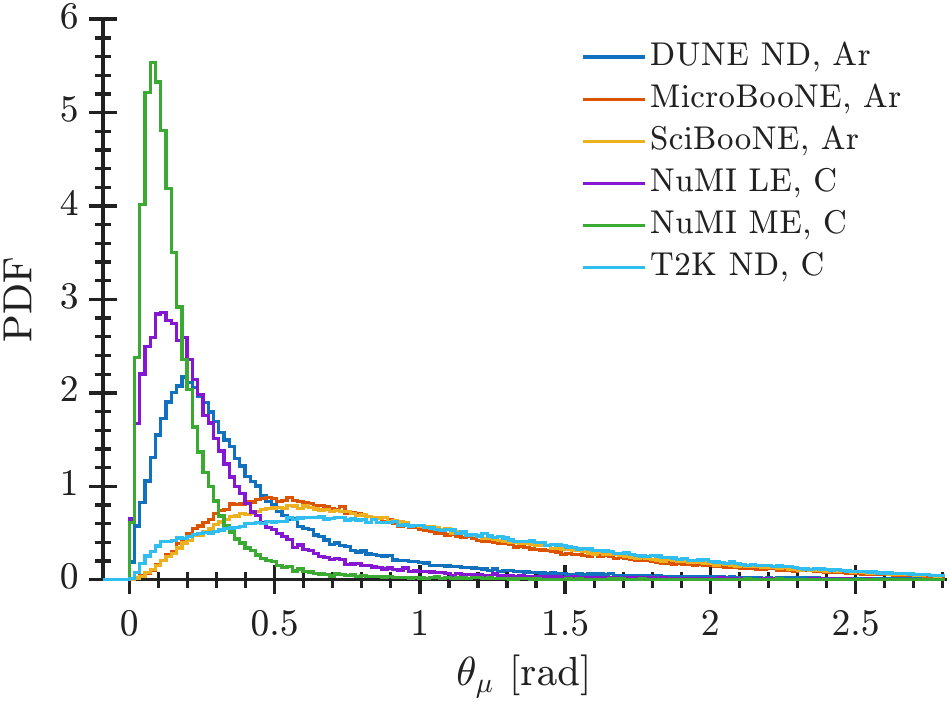}
    \caption{Normalized distributions for several generated flux--target configurations with different neutrino fluxes and target materials: true neutrino energy $E_\nu^{\rm true}$ (top left), outgoing muon energy $E_\mu$ (top right), outgoing proton energy $E_p$ (bottom left), and muon angle $\theta_\mu$ (bottom right).}
    \label{fig:experiemnt_kin}
\end{figure}

To investigate the scaling behavior across different kinematic regimes, we consider several generated flux--target configurations with both on-axis and off-axis neutrino beams. Figure~\ref{fig:experiemnt_kin} shows the normalized distributions of the true neutrino energy \(E_\nu^{\mathrm{true}}\), the outgoing-muon energy \(E_\mu\), the outgoing-proton energy \(E_p\), and the muon angle \(\theta_\mu\) for the configurations considered. The different fluxes substantially modify the standard muon kinematics, including the distributions of \(p_\mu\), \(E_\mu\), \(p_{T\mu}\), and \(\theta_\mu\), by populating different regions of the accessible phase space. These substantially different kinematic distributions provide a
non-trivial test of the stability of \(\psi'\) and the observables
derived from it across the flux--target configurations considered.

The corresponding peak-based estimates are listed in Table~\ref{tab:SQE_peak_compare_exp}. Averaging within each target group gives $\bar S_{\rm LQE}=29~\mathrm{MeV}$ for the carbon configurations and $\bar S_{\rm LQE}=18~\mathrm{MeV}$ for the argon configurations, corresponding to the reference values used in Sec.~\ref{sec:exp_outlook}.
\begin{table}[H]
\centering
\caption{Peak positions of the \(S_{\rm LQE}\) distributions for
the different FHC \(\nu_\mu\) flux--target configurations.
All values are given in MeV.}
\label{tab:SQE_peak_compare_exp}
\begin{tabular}{lc}
\hline
Flux--target configuration & $S_{LQE}^{\rm peak}$ [MeV] \\
\hline
DUNE ND (Ar)                 & 22 \\
MicroBooNE (Ar)              & 16\\
NuMI LE (C)         & 29 \\
NuMI ME (C)         & 31 \\
SciBooNE (Ar)                & 17 \\
T2K ND (C)          & 26 \\
\hline
Average over carbon configurations & 29 \\
Average over Ar configurations         & 18 \\
\hline
\end{tabular}
\end{table}
\FloatBarrier
\section{Analytical expressions for $L_1$, $L_2$, and $L_3$}
\label{app:Ln_analytical}

This appendix gives the analytical expressions for the first three
Taylor coefficients derived from Eq.~\eqref{eq:Ln_def}. The expressions below are written for the neutrino transition \(n\to p\). The antineutrino expressions follow from \(n\leftrightarrow p\). We define
\begin{widetext}
\begin{align}
X_\mu
&\equiv
E_\mu-p_{z\mu}-\widetilde M_n,
&
\widetilde M_n
&\equiv
M_n-\bar S_{\rm LQE},
\label{eq:Xmu_def}
\\
\mathcal D_\mu
&\equiv
\left[
p_{T\mu}^2+\left(M_p+X_\mu\right)^2
\right]
\left[
p_{T\mu}^2+\left(M_p-X_\mu\right)^2
\right].
\label{eq:Dmu_def}
\end{align}

The first-order coefficient is
\begin{equation}
L_1(\vec p_\mu)
=
\frac{\sqrt{2}}{M_n}
\frac{X_\mu^2}{\mathcal D_\mu^{1/2}}.
\label{eq:L1_muon_app}
\end{equation}

The second-order coefficient is
\begin{equation}
L_2(\vec p_\mu)
=
\frac{\sqrt{2}}{M_n^2}
\frac{
X_\mu^3
\left[
X_\mu^3
+\left(M_p^2+p_{T\mu}^2\right)X_\mu
+2M_n\left(M_p^2+p_{T\mu}^2\right)
\right]
}{
\mathcal D_\mu^{3/2}
}.
\label{eq:L2_muon_app}
\end{equation}

The third-order coefficient is
\begin{equation}
L_3(\vec p_\mu)
=
\frac{1}{\sqrt{2}\,M_n^3}
\frac{
X_\mu^4\,\mathcal P_3(X_\mu)
}{
\mathcal D_\mu^{5/2}
},
\label{eq:L3_muon_app}
\end{equation}
where
\begin{align}
\mathcal P_3(X_\mu)
={}&
3X_\mu^6
+2\left(5M_p^2+3p_{T\mu}^2\right)X_\mu^4
+8M_n\left(3M_p^2+p_{T\mu}^2\right)X_\mu^3
\nonumber\\
&+
\left[
3\left(M_p^2+p_{T\mu}^2\right)^2
+8M_n^2\left(M_p^2-p_{T\mu}^2\right)
\right]X_\mu^2
\nonumber\\
&+
8M_n\left(M_p^2+p_{T\mu}^2\right)^2X_\mu
+8M_n^2\left(M_p^2+p_{T\mu}^2\right)^2.
\label{eq:P3_muon}
\end{align}
\end{widetext}
\bibliography{references}
\end{document}